\documentclass[aps,prx,superscriptaddress,reprint]{revtex4-1}
\usepackage{amsmath}
\usepackage{amssymb}
\usepackage{graphicx}
\usepackage[colorlinks,urlcolor=blue]{hyperref}
\usepackage{url}
\usepackage{hyperref}
\usepackage{color,xcolor}
\usepackage{epstopdf}
\usepackage{float}
\usepackage{ulem}
\usepackage[percent]{overpic}
\usepackage{siunitx}

\usepackage[protrusion=true,expansion=true]{microtype}
\begin{document}

\title{Intertwined charge, spin, and orbital degrees of freedom under electronic correlations in the one-dimensional Fe$^{3+}$ chalcogenide chain}
\author{Yang Zhang}
\affiliation{Department of Physics and Astronomy, University of Tennessee, Knoxville, Tennessee 37996, USA}
\affiliation{Materials Science and Technology Division, Oak Ridge National Laboratory, Oak Ridge, Tennessee 37831, USA}
\author{Pontus Laurell}
\affiliation{Department of Physics and Astronomy, University of Missouri, Columbia, Missouri 65211, USA}
\affiliation{Materials Science and Engineering Institute, University of Missouri, Columbia, Missouri 65211, USA}
\author{Gonzalo Alvarez}
\affiliation{Computational Sciences and Engineering Division, Oak Ridge National Laboratory, Oak Ridge, Tennessee 37831, USA}
\author{Adriana Moreo}
\affiliation{Department of Physics and Astronomy, University of Tennessee, Knoxville, Tennessee 37996, USA}
\affiliation{Materials Science and Technology Division, Oak Ridge National Laboratory, Oak Ridge, Tennessee 37831, USA}
\author{Thomas A. Maier}
\affiliation{Computational Sciences and Engineering Division, Oak Ridge National Laboratory, Oak Ridge, Tennessee 37831, USA}
\author{Ling-Fang Lin}
\email{lflin@utk.edu}
\affiliation{Department of Physics and Astronomy, University of Tennessee, Knoxville, Tennessee 37996, USA}
\author{Elbio Dagotto}
\email{edagotto@utk.edu}
\affiliation{Department of Physics and Astronomy, University of Tennessee, Knoxville, Tennessee 37996, USA}
\affiliation{Materials Science and Technology Division, Oak Ridge National Laboratory, Oak Ridge, Tennessee 37831, USA}
\date{\today}

\begin{abstract}
Motivated by recent developments in the study of quasi-one-dimensional iron systems with Fe$^{2+}$, we comprehensively study the Fe$^{3+}$ chalcogenide chain system.
Based on first-principles calculations, the Fe$^{3+}$ chain has a similar electronic structure as discussed before in the iron 2+ chain, due to similar Fe$X_4$ ($X$ = S or Se)
tetrahedron chain geometry. Furthermore, a three-orbital electronic Hubbard model for this chain was constructed by using the density matrix renormalization group method.
A robust antiferromagnetic coupling was unveiled in the chain direction. In addition, in the intermediate electronic correlation $U/W$ region, we found an interesting orbital-selective Mott phase with the coexistence of localized and itinerant electrons ($U$ is the on-site Hubbard repulsion, while $W$ is the electronic bandwidth) based on the orbital-selective behavior observed in the charge fluctuations. Furthermore, we do not observe any obvious pairing tendency in the
Fe$^{3+}$ chain in the electronic correlation $U/W$ region, where superconducting pairing tendencies were reported before in iron ladders. This suggests that superconductivity is unlikely to emerge
in the Fe$^{3+}$ systems. Our results establish with clarity the similarities and differences between Fe$^{2+}$and Fe$^{3+}$ iron chains, as well as iron ladders.
\end{abstract}

\maketitle

\section{Introduction}
The discovery of pressure-induced superconductivity in the two-leg ladder compounds BaFe$_2$S$_3$~\cite{Takahashi:Nm} and BaFe$_2$Se$_3$~\cite{Ying:prb17}, both with electronic density $n = 6$,
has attracted intense interest in the iron chalcogenides with quasi-one-dimensional (Q1D) lattice structure. This discovery has rapidly developed into another exciting new branch of iron-based
superconductors~\cite{Yamauchi:prl15,Zhang:prb17,Zheng:prb18,Wu:prb19,Zhang:prb19,Pizarro:prm,Sun:prb20,Zheng:cp22,Zhang:prbblock,Maier:Prb22}. In those two-leg ladder systems, the interplay among charge, spin, orbital, and lattice degrees of freedom in
a reduced dimensional phase space, induces many interesting phenomena such as orbital-selective magnetism~\cite{Caron:Prb,Caron:Prb12}, spin block states~\cite{Herbrych:prl,Herbrych:block}, ferroelectricity~\cite{Dong:prl14,Aoyama:prb19},
orbital--selective Mott phases (OSMP)~\cite{mourigal:prl15,Patel:osmp,Craco:prb20,Stepanov:prl22,Stepanov:prl}, insulator-metal transitions under pressure~\cite{Zhang:prb18,Materne:prb19}, as well as orbital order~\cite{Takubo:prb17}. Motivated by these intriguing discoveries in two-leg
ladder systems, a natural question arises: can iron chains, with similar Fe$X_4$ ($X$ = S or Se) tetrahedron structures, also display similar physical properties and exotic electronic phases?

\begin{figure*}
\centering
\includegraphics[width=0.96\textwidth]{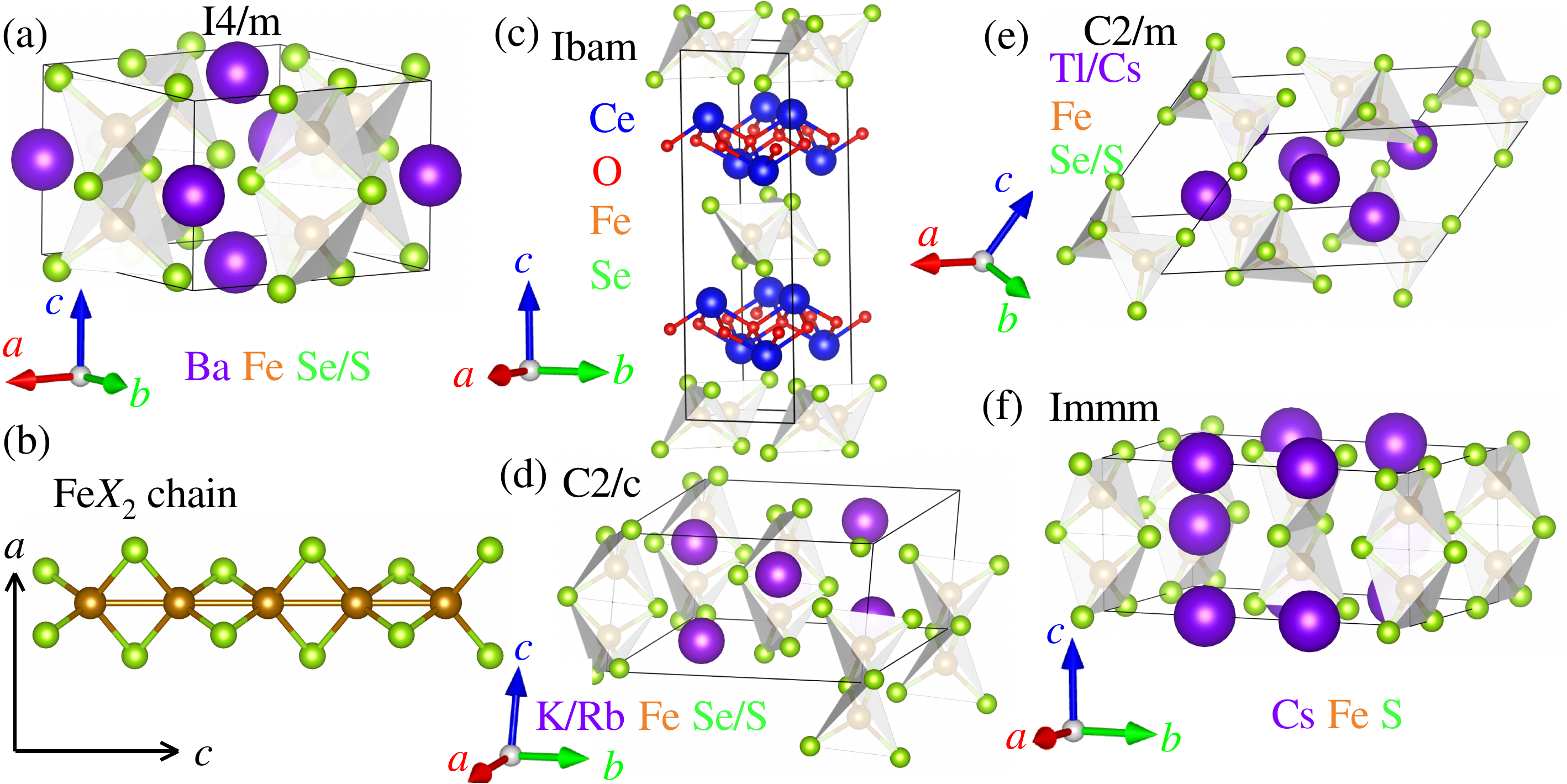}
\caption{ (a) Schematic crystal structure of BaFe$_2$$X$$_4$ ($X$ = S or Se) with space group I4/m (No. 87) in the conventional cell (purple = Ba; brown = Fe; green = Se or S).
(b) Sketch of the $ac$ plane viewed along the iron-chain direction, using BaFe$_2$Se$_4$ as an example. (c) Schematic crystal structure of Ce$_2$O$_2$FeSe$_2$ with space group Ibam (No. 72) in the conventional cell (blue = Ce; red = O; brown = Fe; green = Se). (d) Schematic crystal structure of $A$Fe$X_2$ ($A$ = K or Rb; $X$ = S or Se) with space group C2/c (No. 15) in the conventional cell (purple = K or Rb; brown = Fe; green = Se or S). (e) Schematic crystal structure of TlFe$X_2$ ($X$ = S or Se) and CsFeSe$_2$ with space group C2/m (No. 12) in the conventional cell (purple = Tl; brown = Fe; green = Se or S). (f) Schematic crystal structure of CsFeS$_2$ with space group Immm (No. 71) in the conventional cell (purple = Cs; brown = Fe; green = S. All the crystal structures were visualized with the VESTA code~\cite{momma2011vesta}.}
\label{Crystal}
\end{figure*}

To our best knowledge, unlike the well-discussed iron ladders, the study of iron chains is still in its early stages. Several iron chalcogenide chains, with different electronic densities, have already been synthesized,
such as BaFe$_2$$X_4$ ($X$ = S, Se)~\cite{Berthebaud:jssc,Swinnea:jssc}, Ce$_2$O$_2$FeSe$_2$~\cite{mccabe2011new,mccabe2014magnetism}, and $A$Fe$X$$_2$ ($A$ = K, Rb, Cs and Tl, $X$ = S or Se)~\cite{Bronger:jssc,Seidov:prb01,Seidov:prb16,Li:prb21}.
Their crystal structures are displayed in Fig.~\ref{Crystal}. Although these iron-chain compounds crystallize in different space groups, they all share a common structural character: 1D Fe$X_2$ ($X$ = S or Se) chains built from edge-sharing
Fe$X_4$ ($X$ = S or Se) tetrahedron blocks, as shown in Fig.~\ref{Crystal}(b).

The BaFe$_2$$X_4$ ($X$ = S, Se) family is a typical iron chain with iron 3+~\cite{Berthebaud:jssc,Swinnea:jssc}, corresponding to the electronic configuration of iron $3d^5$, where it has an I4/m crystal structure (No. 87),
as shown in Fig.~\ref{Crystal}(a). A recent neutron experiment revealed a strong antiferromagnetic (AFM) coupling along the chain direction ($c$-axis) and a small FM canting along the $b$ axis~\cite{Liu:prb20}.
Interestingly, the Fe$^{2+}$ material Ce$_2$O$_2$FeSe$_2$ ($3d^6$) has an Ibam crystal structure, which is structurally related to the iron-based superconductor LaFeAsO, as shown in Fig.~\ref{Crystal}(c).
Interestingly, Ce$_2$O$_2$FeSe$_2$ displays a ferromagnetic (FM) coupling along the chain direction~\cite{mccabe2011new,mccabe2014magnetism}. The origin of the FM order along the chains in Ce$_2$O$_2$FeSe$_2$ can be explained by a novel
``half-full'' mechanism involving the large entanglement between half-filled and fully occupied orbitals proposed by L-F. Lin et. al.~\cite{Lin:prl}. In addition, an interesting FM OSMP with both localized and itinerant electrons has also
been discussed with electronic correlations under crystal-field splitting and doping effects~\cite{LindopingOSMP1,LindopingOSMP2} by using the many-body density matrix renormalization
group (DMRG) technique.

\begin{table*}
\centering\caption{Summary of key structural and magnetic properties of various iron chain systems, including space groups, valences of iron, nearest-neighbor Fe-Fe distances (\AA), Fe-$X$ ($X$ = S or Se) bond lengths, magnetic correlations along the iron chain,
magnetic transition temperatures $T^*$ (K) and magnetic moments ($\mu_{\rm B}$/Fe), as obtained from experimental studies. }
\begin{tabular*}{0.96\textwidth}{@{\extracolsep{\fill}}lllllllc}
\hline
\hline
Column  & Space group & Valence & Fe-Fe  & Fe-$X$  & Magnetism & $T^*$ & Magnetic moment\\
\hline
BaFe$_2$Se$_4$~\cite{Berthebaud:jssc,Liu:prb20}    & I4/m & +3 & 2.742 & 2.349 & AFM & 310 & 2.09\\
BaFe$_2$S$_4$~\cite{Swinnea:jssc}   & I4/m  & +3 & 2.646 & 2.218 & -- & -- & --\\
Ce$_2$O$_2$FeSe$_2$~\cite{mccabe2011new,mccabe2014magnetism}   & Ibam & +2 & 2.850 & 2.447 & FM & 171 & 3.33\\
KFeSe$_2$~\cite{Bronger:jssc,Seidov:prb16} & C2/c & +3 & 2.815 & 2.363/2.369 & AFM & 310 & 3\\
KFeS$_2$~\cite{Bronger:jssc,Seidov:prb16} & C2/c & +3 & 2.698 & 2.231/2.237 & AFM & 250 & 2.43 \\
RbFeSe$_2$~\cite{Bronger:jssc,Seidov:prb16} & C2/c & +3 & 2.831 & 2.383/2.386 & AFM & 250 & 2.66\\
RbFeS$_2$~\cite{Bronger:jssc,Seidov:prb16} & C2/c & +3 & 2.716 & 2.235/2.245 & AFM & 188 & 1.83 \\
CsFeSe$_2$~\cite{CsFeSe2} & C2/m & +3 & 2.806/2.838 & 2.356/2.366 & -- & -- & --\\
TlFeSe$_2$~\cite{Klepp:1979,Seidov:prb01} & C2/m & +3 & 2.737/2.753 & 2.344/2.355 & AFM & 290 & 2.1\\
TlFeS$_2$~\cite{Klepp:1979,Seidov:prb01} & C2/m & +3 & 2.641/2.663 & 2.257/2.295 & AFM & 196 & 1.85 \\
CsFeS$_2$~\cite{Ho:1985,Seidov:prb16} & Immm & +3 & 2.699/2.721 & 2.262/2.220 & AFM & -- & 1.88 \\
\hline
\end{tabular*}
\label{Table1}
\end{table*}

The $A$Fe$X_2$ ($A$ = K, Rb, Cs and Tl, $X$ = S or Se) family is also an Fe$^{3+}$ chain system, with electronic density $3d^5$. They have the same chemical formula but different space groups~\cite{Bronger:jssc,CsFeSe2,Klepp:1979,Ho:1985},
as displayed in Figs.~\ref{Crystal}(d-f). Neutron diffraction experiments also found a strong AFM coupling along the chain direction~\cite{Bronger:jssc,Seidov:prb16} in those iron 3+ chains. As summarized in Table~\ref{Table1},
both Fe$^{2+}$ ($3d^6$) and Fe$^{3+}$ ($3d^5$) chain systems exhibit very similar crystal structures along the chain direction, suggesting potential similarities in their physical properties. But they also have some differences. Considering
the present studies on iron Q1D systems, several physical questions naturally arise. At intermediate couplings, could the OSMP regime be obtained in the iron 3+ chain? What causes the different
magnetic coupling along the chain direction for the
Fe$^{3+}$ and Fe$^{2+}$ chains? Can a unified picture be established to understand magnetic correlations across different iron chain systems?  Can superconductivity happen in the Fe$^{3+}$ chain?

To address these questions,  we comprehensively study the Fe$^{3+}$ chalcogenide chain, using BaFe$_2$Se$_4$ as a model system, employing a combination of density functional theory (DFT) and density matrix renormalization group (DMRG) methods.
Our DFT results show that the five iron orbitals of BaFe$_2$Se$_4$ also display entangled bonding and antibonding characteristics between different orbitals with large interorbital hoppings,
which is quite similar to the case discussed in the iron 2+ chain~\cite{Lin:prl}. The three-orbital model DMRG calculations indicate a very robust AFM coupling along the iron chain in the Fe$^{3+}$ system. At the intermediate electronic correlation $U/W$ region,
an interesting OSMP regime was also obtained in the Fe$^{3+}$ chain with the coexistence of localized and itinerant electrons. By modifying the value of electronic density to $n = 4$ in our three-orbital electronic Hubbard model, corresponding to iron 2+,
using the hopping matrix and crystal field splitting from our case, remarkably we found the FM state. In addition, changing the hopping matrix and crystal field splitting to those of the Fe$^{2+}$ chain, using the electronic density $n = 3$ (corresponding
to the iron 3+), we obtained the AFM state. Those results supported the novel mechanism proposed by our previous study~\cite{Lin:prl}, indicating that the electronic density of iron is responsible for the different magnetic behaviors in iron 3+ and 2+
chains. Moreover, we do not observe a clear pairing tendency in the Fe$^{3+}$ chain at both intermediate and strong electronic correlations, suggesting that superconductivity is unlikely to emerge in these systems, or that the superconducting transition temperature, if present, would be very low. In this case, our results well establish the similarities and differences between Fe$^{2+}$ and Fe$^{3+}$ iron chains, in contrast with iron Fe$^{2+}$ ladders.

\section{Methods}

\subsection{DFT method}
To calculate the electronic structure of BaFe$_2$Se$_4$, first-principles DFT calculations were performed using the Vienna {\it ab initio} simulation package (VASP),
within the projector augmented wave (PAW) method~\cite{Kresse:Prb,Kresse:Prb96,Blochl:Prb}. The electronic correlations were considered by using the generalized gradient approximation (GGA)
within the Perdew-Burke-Ernzerhof exchange correlation potential~\cite{Perdew:Prl}. Here, the plane-wave cutoff was set as $500$ eV, and the $k$-point mesh was $8\times8\times12$ for the non-magnetic calculations.
For the calculation of the density-of-states the $k$-point mesh used was to $12\times12\times16$. In addition to the standard DFT calculations, the maximally localized Wannier functions (MLWFs) method was employed
to fit the Fe $3d$'s bands to obtain hopping parameters and crystal field splitting by using the WANNIER90 packages~\cite{mostofi:cpc}.

\subsection{DMRG method}
To study the three-orbital electronic Hubbard model discussed in Sec. III [See Eqs.~(\ref{eq1} and \ref{eq2})], we used the many-body DMRG method~\cite{white:prl,white:prb,schollwock2005density}, where the DMRG++ computer package was employed~\cite{alvarez2009density}.
In the present DMRG calculations, at least $1400$ states were kept and up to $21$ sweeps were performed during the finite-size algorithm evolution. In previous studies of the same three-orbital Hubbard model but for materials with different hoppings and crystal-field splittings, using the same value for the number of kept states was shown to provide a good description for the chain systems studied, by considering the balance of physical quantities accuracy and computational costs~\cite{Rincon:prl14,Lin:prl,zhang:prb21}. In addition, an electronic filling $n = 3$
in the three orbitals was considered. Here, a cluster chain lattice geometry with open-boundary conditions (OBC) was used for different lengths $L$. The truncation error remains below $10^{-6}$
for the calculation of spin-spin correlation, spin structure factor, charge occupancy, charge fluctuation, and the squared spin. Furthermore, different lengths $L$ and states $m$ were considered in the study of the pairing tendencies.

To help the readers reproduce our results, an example input file and additional details of input parameters are  described in the Supplemental Materials~\cite{Supplemental}.

\section{Model system}

\subsection{DFT results of BaFe$_2$Se$_4$}
First, we briefly present the DFT results for the representative material BaFe$_2$Se$_4$. Based on the experimental crystal structure~\cite{Berthebaud:jssc}, we investigated the electronic structures for the nonmagnetic state
of BaFe$_2$Se$_4$. The DFT results clearly establish both the similarities and differences between the $n = 5$ and $n = 6$ iron chain systems, arising from their structurally similar chain geometry.

\begin{figure}
\centering
\includegraphics[width=0.48\textwidth]{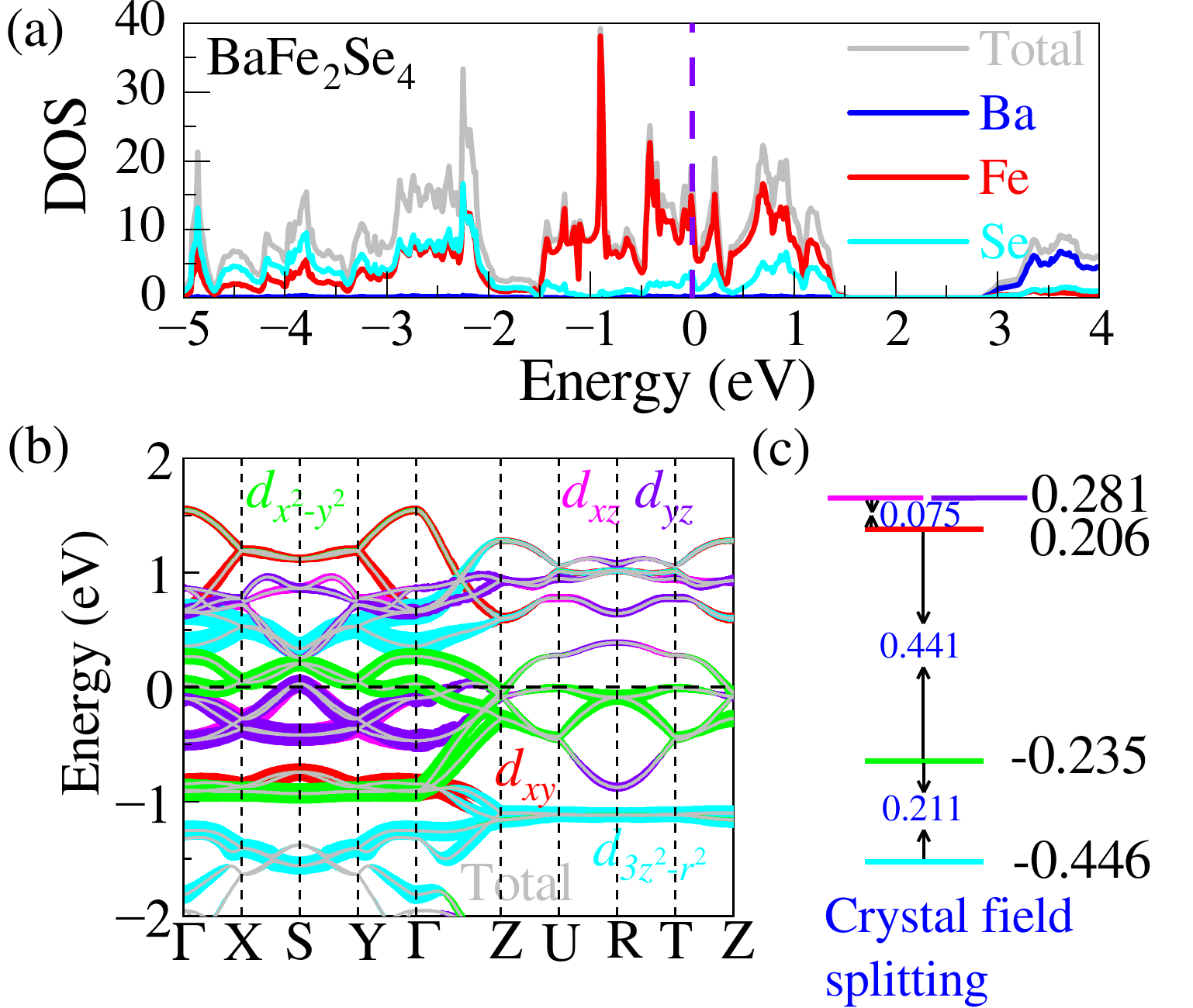}
\caption{ (a) Density of states near the Fermi level of BaFe$_2$Se$_4$ for the non-magnetic phase (Gray = Total; blue = Ba; red = Fe; cyan = Se). (b) Projected band structures of the nonmagnetic phase
of BaFe$_2$Se$_4$. Note that the local \{$x$, $y$, $z$\} axes of projected orbitals correspond to crystal axes \{$a$, $b$, $c$\}, where the $z$ axis is the $c$ axis and the $x$ or $y$ axes are the
$a$ or $b$ axes. The weight of each Fe orbital is represented by the size of the (barely visible) circles. (c) The crystal splitting energies of five Fe $3d$ orbitals obtained by Wannier fitting. More details
can be found in the Supplemental Materials.}
\label{DFT}
\end{figure}

As shown in Fig.~\ref{DFT}(a), the states near the Fermi level are mainly contributed by the Fe $3d$ orbitals hybridized with the Se $4p$ state.
Furthermore, the Fe $3d$ bands of BaFe$_2$Se$_4$ are mainly located in the range of energy from $-1.5$ to $1.5$ eV, while the Se $4p$ bands are extended in real space.
Compared with the bandwidth of iron $3d$ orbitals of the $n = 6$ chain Ce$_2$O$_2$FeSe$_2$ system ($\sim 2.5$ eV)~\cite{Lin:prl}, the Fe $3d$ bandwidth is slightly
larger ($\sim 3$ eV) for this $n = 5$ iron chain, indicating a slightly enhanced nature of the itinerant behavior of $3d$ states in BaFe$_2$Se$_4$.

Furthermore, the reported spin magnetic moments for the  Fe$^{3+}$ chains range from approximately 1.8 to 3 $\mu_{\rm B}$/Fe, as displayed in Table~\ref{Table1}, which is significantly lower than the value of the spin magnetic moment
of the free ions Fe$^{3+}$ ($\sim 5$ $\mu_{\rm B}$/Fe). This behavior is quite common in the iron-based superconductors with Fe$X_4$  tetrahedral coordination~\cite{Dagotto:Rmp,Dai:Rmp,Ootsuki:prb15}. Using the local spin density approximation (LSDA), which is
widely applied in the DFT context of the Q1D and two-dimensional iron-based chalcogenides systems~\cite{Zhang:prb18,Zheng:prb18,Dagotto:Rmp,Dai:np}, we calculated that
the local magnetic moment of Fe is about $2.82$ $\mu_{\rm B}$/Fe. This value is
also larger than the experimental values ($\sim 2.09$ $\mu_{\rm B}$/Fe). This may be caused by the coexistence of localized Fe spins and itinerant electrons in the Q1D systems~\cite{Ootsuki:prb15}. Another possible reason is
that the quantum fluctuations are strong for the iron-based Q1D systems~\cite{Dagotto:Rmp}.

The DFT band structure of the nonmagnetic phase of BaFe$_2$Se$_4$ [see Fig.~\ref{DFT}(b)] is more dispersive from $\Gamma$ to $Z$ (parallel to the chains) along the chains than along other directions,
such as $\Gamma$ to $X$ along the $a$-axis. This anisotropy indicates a dominant QlD behavior along the $k_z$ axis caused by the Q1D iron chain lattice geometry. Similar to the case of Ce$_2$O$_2$FeSe$_2$ with $n = 6$, when
interactions are neglected, the five iron orbitals of BaFe$_2$Se$_4$ also display entangled bonding and antibonding characteristics. In addition, we also obtained crystal splitting energies
for five iron $3d$ orbitals~\cite{Supplemental}, based on the
maximally localized Wannier functions~\cite{mostofi:cpc}, as displayed in Fig.~\ref{DFT}(c), which is similar to the case of iron 2+ chain Ce$_2$O$_2$FeSe$_2$~\cite{Lin:prl}. Due to the
similar FeSe$_2$ chain geometry in the Fe$^{2+}$
and Fe$^{3+}$ chains, the interorbital hoppings (nonzero off-diagonal matrix elements) are also found to be large in the case of BaFe$_2$Se$_4$ ($n = 5$). More results of Wannier fitting can be found in the Supplemental Material.

\subsection{Model Hamiltonian}
In the present study, as an example of a Fe$^{3+}$ chalcogenide chain, we employ a canonical three-orbital Hubbard model defined on a 1D chain geometry,
including the kinetic energy and interaction terms given by $H = H_k + H_{int}$. The tight-binding kinetic component is:
\begin{eqnarray}
\label{eq1}
H_k = \sum_{\substack{i\sigma\gamma\gamma'}}t_{\gamma\gamma'}(c^{\dagger}_{i\sigma\gamma}c^{\phantom\dagger}_{i+1\sigma\gamma'}+H.c.)+ \sum_{i\gamma\sigma} \Delta_{\gamma} n_{i\gamma\sigma},
\end{eqnarray}
where the first term represents the hopping of an electron between orbitals $\gamma$ at site $i$ and orbitals $\gamma'$ at the nearest neighbor (NN) site $i+1$. $c^{\dagger}_{i\sigma\gamma}$($c^{\phantom\dagger}_{i\sigma\gamma}$) is the standard creation (annihilation) operator, $\gamma$ and $\gamma'$ indicate the different orbitals, and $\sigma$ is the $z$-axis spin projection. $\Delta_{\gamma}$ are the crystal-field splittings of different orbitals $\gamma$.

The electronic interaction portion of the Hamiltonian includes the standard intraorbital Hubbard repulsion $U$, the electronic repulsion  $U'$ between electrons
at different orbitals, the Hund's coupling $J_H$, and the on-site inter-orbital electron-pair hopping terms. Formally, it is written as:
\begin{eqnarray}\label{eq2}
H_{int}= U\sum_{i\gamma}n_{i \uparrow\gamma} n_{i \downarrow\gamma} +(U'-\frac{J_H}{2})\sum_{\substack{i\\\gamma < \gamma'}} n_{i \gamma} n_{i\gamma'} \nonumber \\
-2J_H  \sum_{\substack{i\\\gamma < \gamma'}} {{\bf S}_{i,\gamma}}\cdot{{\bf S}_{i,\gamma'}}+J_H  \sum_{\substack{i\\\gamma < \gamma'}} (P^{\dagger}_{i\gamma} P_{i\gamma'}+H.c.),
\end{eqnarray}
where the standard relation $U'=U-2J_H$ is assumed and the interorbital pair hopping is $P_{i\gamma}$=$c_{i \downarrow \gamma} c_{i \uparrow \gamma}$.

Here, we consider a three-orbital Hubbard model with three electrons per site (more details can be found in the Method section), corresponding to the iron valence Fe$^{3+}$ ($3d^5$). In principle, for the iron 3+ chain materials with $3d^5$ configuration, all five iron $3d$ orbitals contribute to low-energy bands near the Fermi surface [see Fig.~\ref{DFT}(b)]. However, in the DMRG calculations, five-orbital Hubbard-type models require significantly larger computing costs with more hoppings, crystal-field splittings, Hund's coupling, and Hubbard interactions due to the exponential increase of the local Hilbert space and substantial entanglement growth, which is impossible to do with our present computing resources. Thus, here we considered the three-orbital Hubbard model in our DMRG simulations, due to both physical and computational considerations, by following our previous study on the iron chains and ladders with a similar edge-sharing Fe$X_4$ tetrahedron structure~\cite{Lin:prl,Rincon:prl14,Herbrych:nc18}. The case $n = 4$ (four electrons in three orbitals), the typical electronic density used in iron superconductors where iron is in a valence Fe2+, corresponds to six electrons in five orbitals~\cite{Daghofer:prb10,Rincon:prl14,Herbrych:nc18}. Thus, the Fe$^{3+}$ chain discussed here corresponds to $n = 3$ in the three-orbital model. Note that the three-orbital model was already widely employed in theoretical studies of real iron superconductor systems, and multiple efforts provided evidence that the three-orbital approximation gives a good description of the material's physical properties~\cite{Patel:osmp,Lin:prl,Luo:prb10,Daghofer:prb10,Herbrych:nc18}. For example, in the Fe$^{2+}$ iron ladder system BaFe$_2$Se$_3$, the three-orbital DMRG studies~\cite{Herbrych:nc18,Patel:osmp} successfully captured the essential physics of the block orbital-selective Mott phase, providing a consistent description of the original experimental discoveries~\cite{Caron:Prb,Caron:Prb12}. A similar situation arises in iron chain systems. Our previous three-orbital DMRG work~\cite{Lin:prl} on the Fe$^{2+}$ chain compound Ce$_2$O$_2$FeSe$_2$ demonstrated excellent agreement between theory and experiment~\cite{mccabe2014magnetism}. Furthermore, for the Fe$^{2.5+}$ iron chain K$_3$Fe$_2$Se$_4$, our three-orbital model successfully reproduces the block-type magnetic order observed in neutron scattering experiments~\cite{Gao:prl}. Taken together, these examples support the conclusion that the three-orbital model captures the essential physics of Fe-based chain and ladder systems.

The crystal-field splitting and  hopping matrix elements are obtained from BaFe$_2$Se$_4$, as a concrete example. The non-interacting tight-binding band structure along the chain direction, using the nearest-neighbor hopping and crystal-field splitting obtained from BaFe$_2$Se$_4$, is displayed in Fig.~\ref{DMRG}(a), which qualitatively reproduces the corresponding DFT bands.

The hopping matrix for the three-orbital chain system is obtained from MLWFs of BaFe$_2$Se$_4$ in orbital space as follows:

\begin{equation}
\begin{split}
t_{\gamma\gamma'} =
\begin{bmatrix}
         -0.457     &   0.045	   &      -0.495	   	       \\
          0.045	    &   0.224	   &       0.107	   	       \\
         -0.495	    &   0.107	   &      -0.439  	
\end{bmatrix}.\\
\end{split}
\end{equation}
The crystal field splitting of the three orbitals is chosen as $\Delta_{0} = -0.446$, $\Delta_{1} = -0.235$, and $\Delta_{2} = 0.206$, all in eV units.  The total kinetic energy bandwidth $W$ is about $2.6$ eV
for the three-orbital model. All parameters mentioned above, including the hopping matrix and crystal fields, were discussed in ``Supplemental Note II''~\cite{Supplemental}.

\subsection{Observables}
To study the three-orbital 1D Hubbard model varying $U/W$, several observables were measured using the DMRG many-body technique. Although we use a SU(2)-symmetric Hamiltonian model as defined in this work, we did not explicitly implement the SU(2) symmetry in the DMRG simulations. Instead, we implemented the canonical DMRG with U(1) spin symmetry only, conserving the total $S^z$. It is well known that DMRG calculations only based on U(1), rather than full SU(2), can describe ferromagnetic tendencies correctly. Within this framework, spin-up and spin-down sectors are treated explicitly on a U(1) basis. Accordingly, the number of kept states reported in this work refers to U(1)-symmetric bases, rather than SU(2) multiplet bases.

The spin-spin correlation is defined as:
\begin{eqnarray}
S(r)=\langle{{\bf S}_i \cdot {\bf S}_j}\rangle.
\end{eqnarray}
Here $r=\left|{i-j}\right|$, and the spin at site $i$ is
\begin{eqnarray}
{\bf S}_i = \frac{1}{2}\sum_\gamma\sum_{\alpha\beta}c^\dagger_{i\gamma\alpha}{\bf \sigma}_{\alpha\beta}c^{\phantom\dagger}_{i\gamma\beta}\,,
\end{eqnarray}
where ${\bf \sigma}_{\alpha\beta}$ are the matrix elements of the Pauli matrices.

The corresponding structure factor for spin is:
\begin{eqnarray}
S(q)=\frac{1}{L}\sum_{\substack{j,k}}e^{-iq(j-k)}\langle {\bf{S}}_{k}\cdot {\bf{S}}_{j}\rangle,
\end{eqnarray}

The site-average occupancy of orbitals is:
\begin{eqnarray}
n_{\gamma}=\frac{1}{L}{\langle}n_{i\gamma\sigma}\rangle.
\end{eqnarray}

The orbital-resolved charge fluctuation is defined as:
\begin{eqnarray}
\delta{n_{\gamma}}=\frac{1}{L}\sum_{i}({\langle}n_{\gamma,i}^2\rangle-{\langle}n_{\gamma,i}{\rangle}^2).
\end{eqnarray}

The mean value of the squared spin for each orbital is defined as:
\begin{eqnarray}
\langle {\bf{S}}^2\rangle_{\gamma} =\frac{1}{L}\sum_{\substack{i}} \langle {\bf{S}}_{i,\gamma}\cdot{\bf{S}}_{i,\gamma}\rangle.
\end{eqnarray}

While it is true that in 1D or Q1D systems, alternative non-magnetic ordered phases, such as dimerized states, valence bond solid phases, and charge-density-wave states, can in principle emerge in certain parameter regimes, their existence should have appeared in our calculations. But they did not. Maybe by changing some model parameters, these competing phases may become relevant in some regions. But we employ hoppings and crystal fields from DFT for the materials of our focus, and then solve the Kanamori Hubbard multiorbital model within our best abilities via DMRG. Playing with the vast parameter space to discover additional phases is charming as an idea, but totally beyond the scope of the present work that focuses on specific materials. DFT renders our work realistic for the materials of our focus.

\section{DMRG calculations}

\subsection{Spin-spin correlation}

\begin{figure}
\centering
\includegraphics[width=0.48\textwidth]{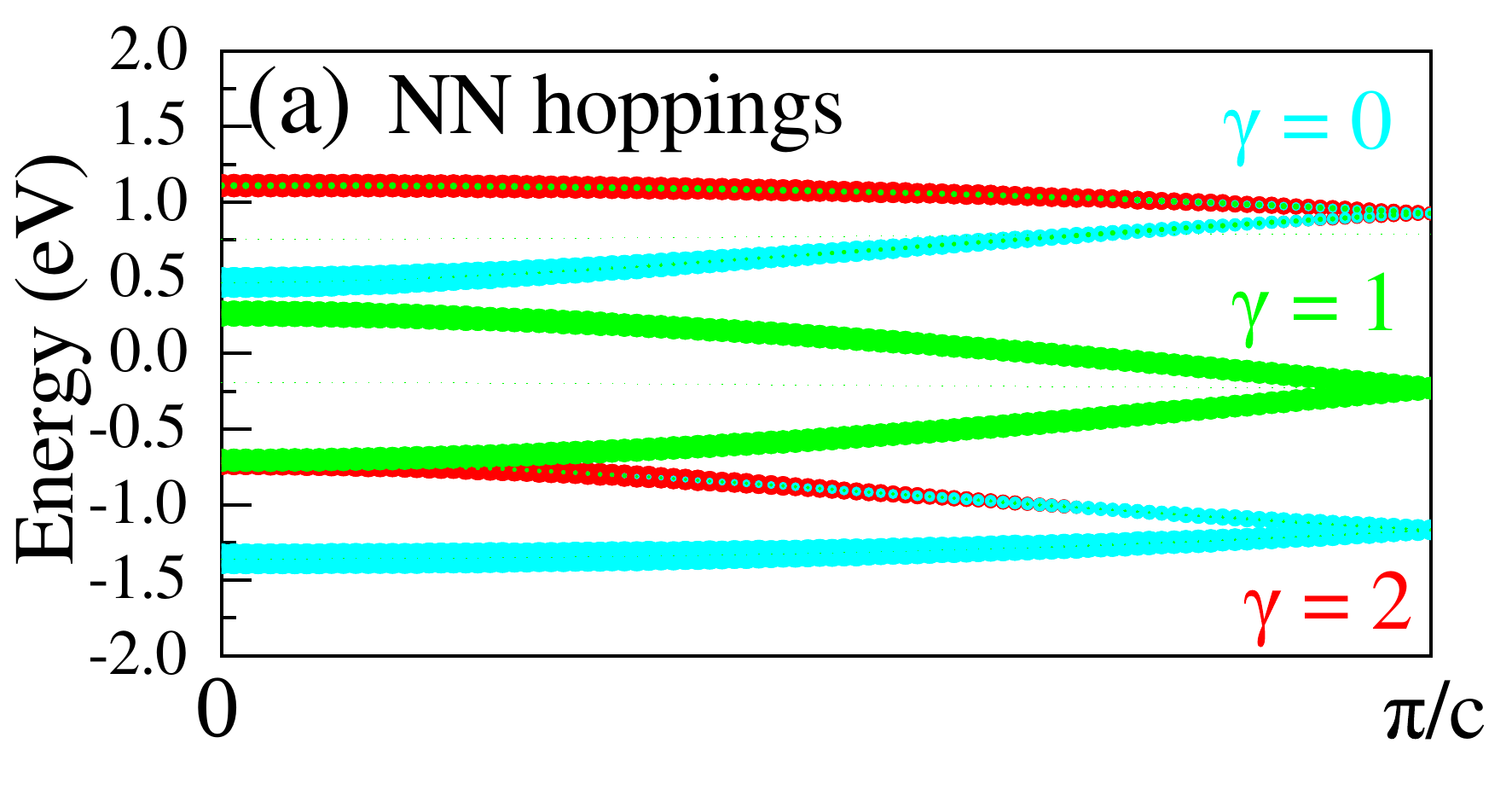}
\includegraphics[width=0.48\textwidth]{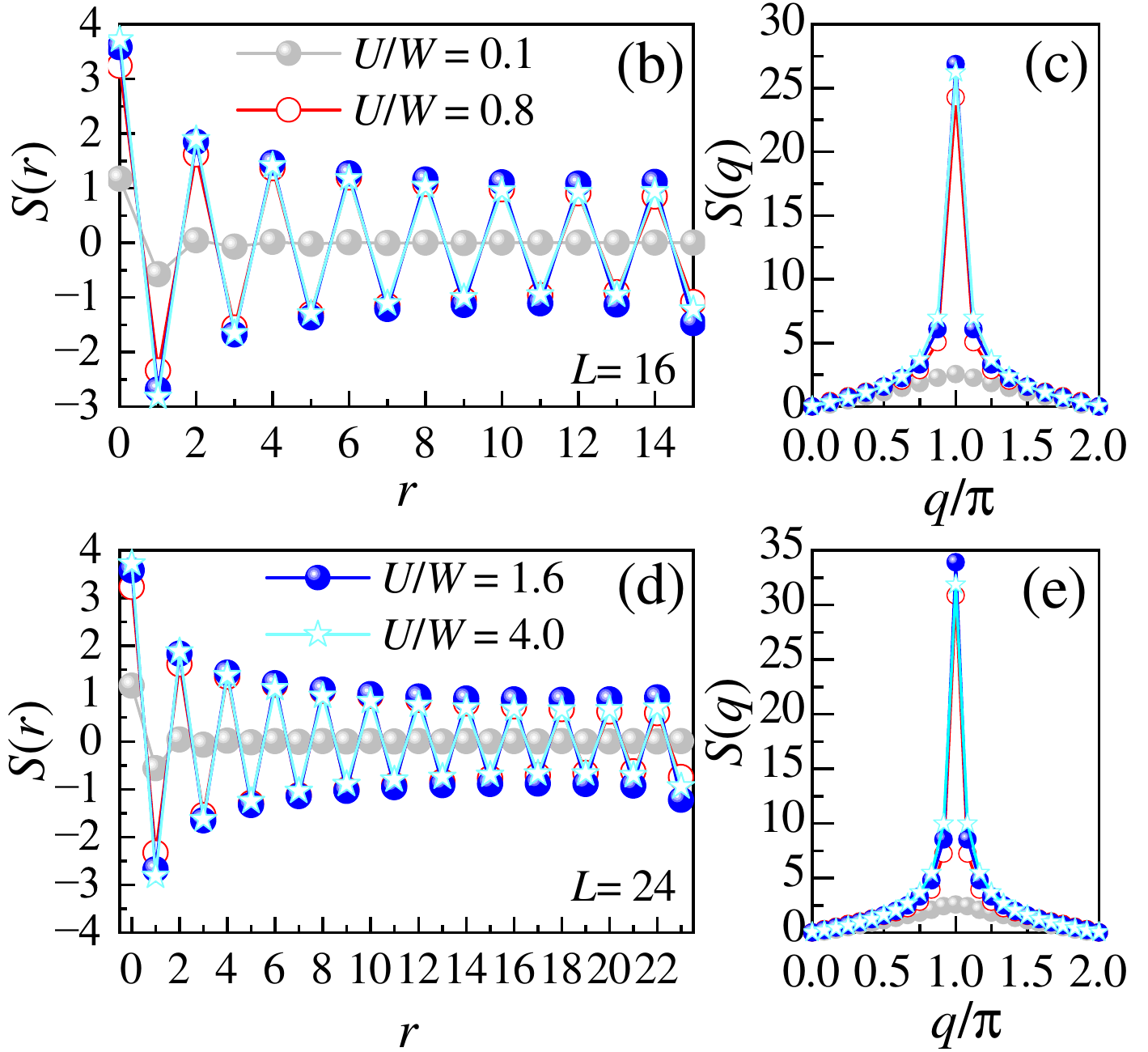}
\includegraphics[width=0.48\textwidth]{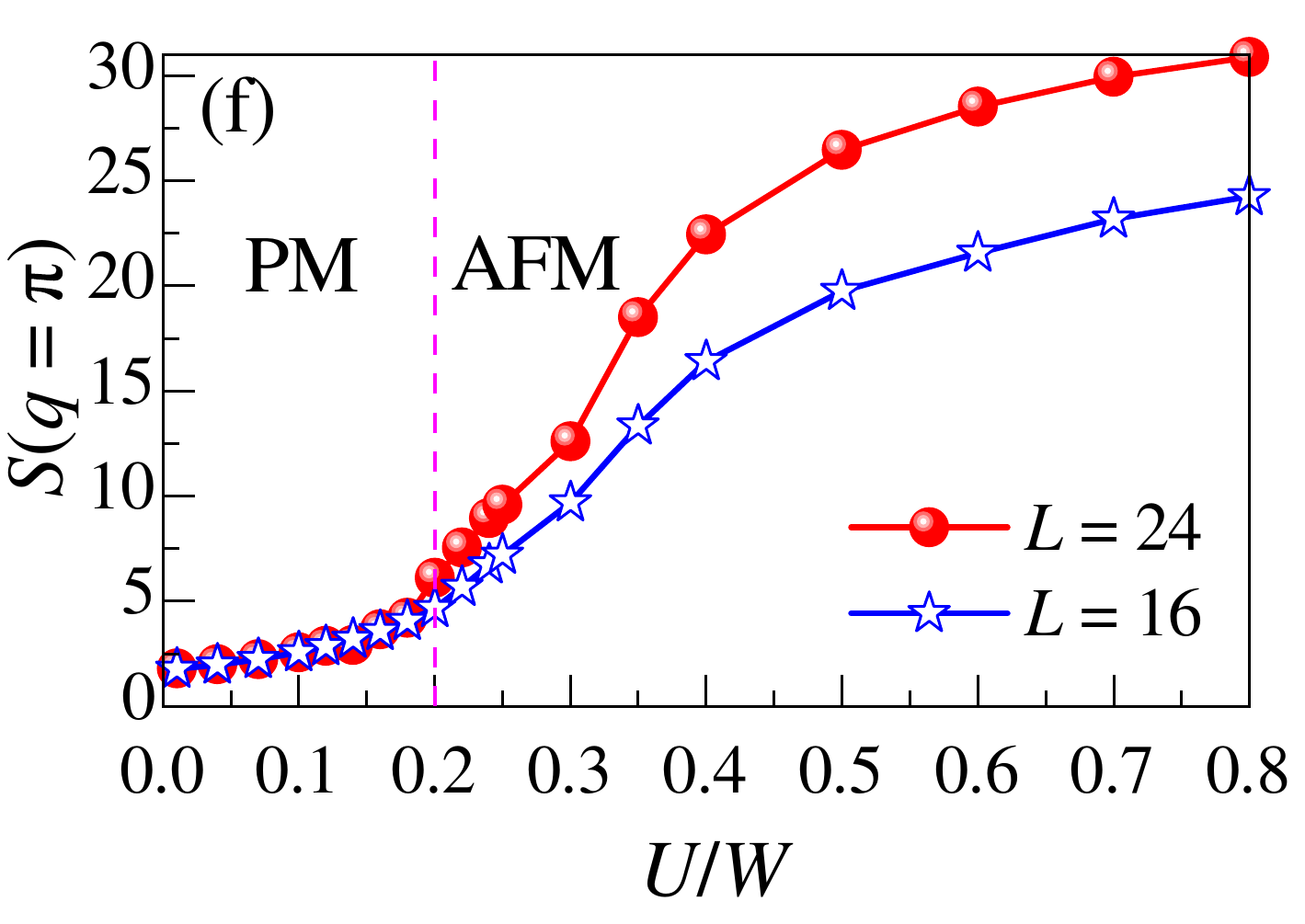}
\caption{(a) The non-interacting tight-binding band structure along the chain direction with nearest-neighbor hopping obtained from example material BaFe$_2$Se$_4$. In this particular material (BaFe$_2$Se$_4$), there are
two iron atoms in the chain, thus there are six bands in our tight-binding calculations. (b) and (d) Spin-spin correlation $S(r)=\langle{{\bf S}_i \cdot {\bf S}_j}\rangle$ (with $r=\left|{i-j}\right|$ in real space). (c) and (e) the spin structure factor $S(q)$, as a function of $U/W$, at $J_H$/$U$ = 0.25, respectively. Here, we use a chain lattice geometry with $L = 16$ and $L = 24$, respectively. (f) Spin structure factor $S(q = \pi)$ for different electronic correlations $U/W$, and at different lengths $L = 16$ and $L = 24$, respectively.}
\label{DMRG}
\end{figure}

As shown in Fig.~\ref{DMRG}(b), the spin-spin correlation $S(r)$=$\langle{{\bf S}_i \cdot {\bf S}_j}\rangle$ decays rapidly as distance $r$ increases in the small Hubbard interaction $U/W$ region (see $U/W = 0.1$ as an example), where the distance is defined as $r=\left|{i-j}\right|$, with $i$ and $j$ site indices~\cite{Spin-spin}. Accordingly, the spin structure factor $S(q)$ does not show an
obvious peak in reciprocal space, indicating a paramagnetic (PM) behavior. In this region, the kinetic term plays the leading role, resulting in a metallic PM state.

As $U/W$ increases, the spin-spin correlation $S(r)$ increases at larger distances (see the $U/W = 0.8$), as displayed in Fig.~\ref{DMRG}(b). In addition, the spin structure factor $S(q)$ also
displays a sharp peak at $q = \pi$, indicating staggered spin pattern (see Fig.~\ref{DMRG}(c)). Note that eventually at large distances quantum fluctuations will prevent full long-range order in the one
dimensional model. Strictly in one-dimensional systems with continuous symmetry, as implied by the Lieb-Schultz-Mattis theorem, the spin-spin correlations eventually should decay slowly as a power law. In our results, although the strong spin correlations unveiled here resemble long-range order, in much larger systems quite likely the power-law decay will be observed. Moreover, we note that weak couplings between chains will stabilize the unveiled tendencies into long-range patterns with the canonical staggered antiferromagnetic order for the real material BaFe$_2$Se$_4$ system. Here, we do not at all make any claim of long-range orders in one dimension in the paper. We focus on the {\it dominant tendencies}, which likely will become true long-range order after adding weak coupling between chains. This staggered spin pattern is also supported by DFT+$U$ calculations for BaFe$_2$Se$_4$, as discussed in Supplementary Materials V~\cite{Supplemental}.

By continuing increasing $U/W$, we did not find any other magnetic order in the range of $U/W$ we studied ($U/W \leq 10$), indicating the AFM coupling is robust along the chain direction (see  Figs.~\ref{DMRG}(b) and (c)).
Furthermore, we also calculated $S(r)$ and $S(q)$ at larger $L$ ($L = 24$), as displayed in Figs.~\ref{DMRG}(d) and (e), respectively. The results are similar to the results of $L = 16$. Then, we conclude that the
staggered AFM is robust in the $n = 5$ iron chain against changes of lattice length $L$. This conclusion is in good agreement with the present experimental results for $n = 5$ iron chain systems:
BaFe$_2$Se$_4$~\cite{Liu:prb20}, $A$Fe$X_2$ ($A$ = K or Rb and $X$ = S or Se)~\cite{Bronger:jssc}, and TlFe$X$$_2$ ($X$ = S or Se)~\cite{Seidov:prb01}. This strong AFM coupling chain can be easily understood.
In the half-filled multi-orbital model, both intraorbital and interorbital hoppings favor the AFM coupling due to the Pauli principle.

To better understand the PM-AFM phase transition in this system, we calculated the spin structure factor at $q = \pi$ for different values of $U/W$. Fig.~\ref{DMRG}(f) clearly indicates an PM-AFM phase transition around $\sim$ $U/W = 0.2$ with a peak appearing in the spin structure factor $S(q)$, where the $S(q = \pi)$ begins to rapidly increase, indicating that AFM order forms. Furthermore, it
also should be noticed that the specific boundary values of the PM-AFM phase transition is only a crude approximation, because finite lattice size effects and the use of a limited number of states in DMRG would affect the specific boundary values. In the present work, our purpose is not to do a rigorous determination of the boundary of AFM correlations, but to determine the existence of the dominant different phase regions.

\subsection{Charge fluctuations}
\begin{figure}
\centering
\includegraphics[width=0.48\textwidth]{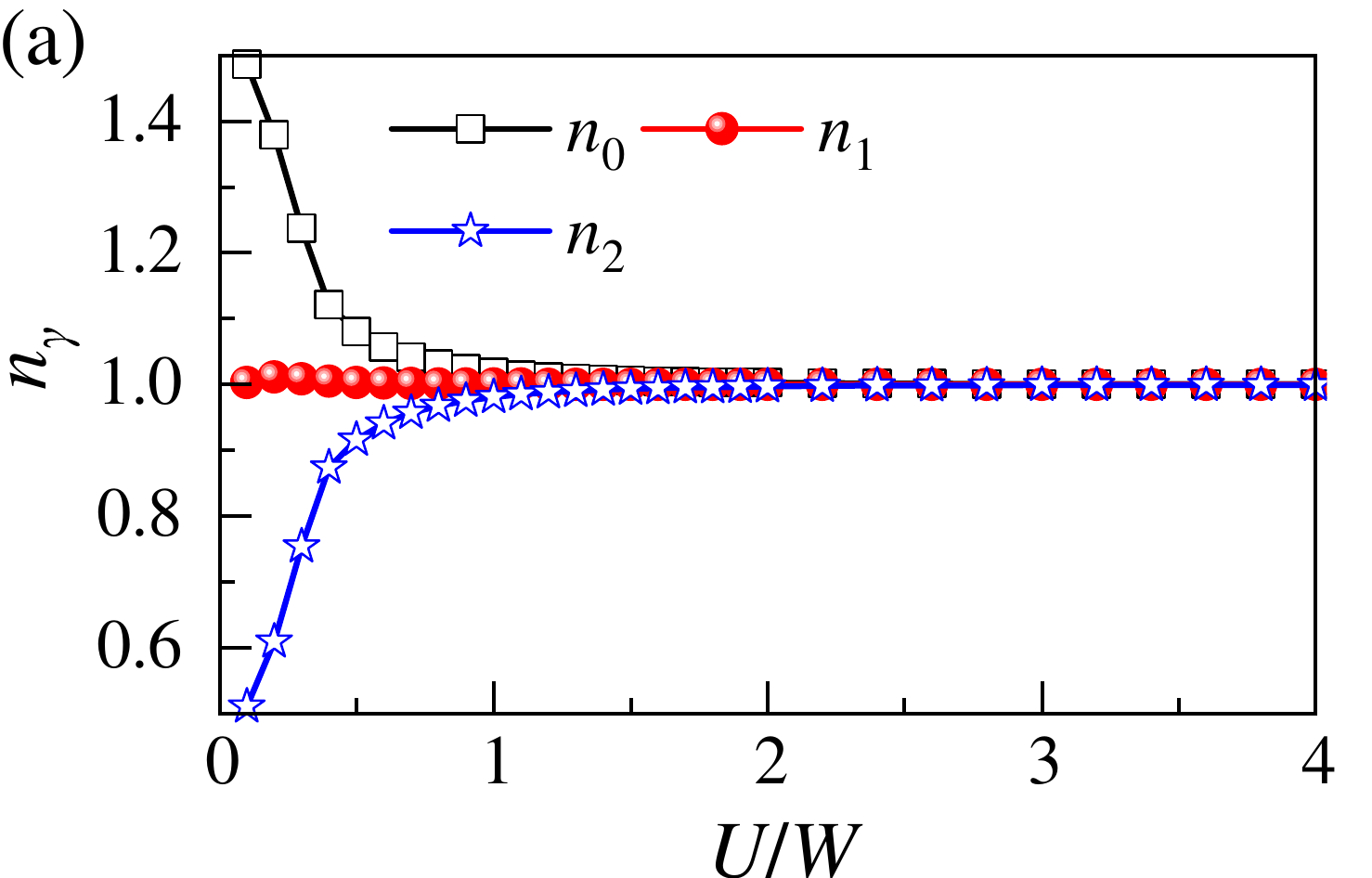}
\includegraphics[width=0.48\textwidth]{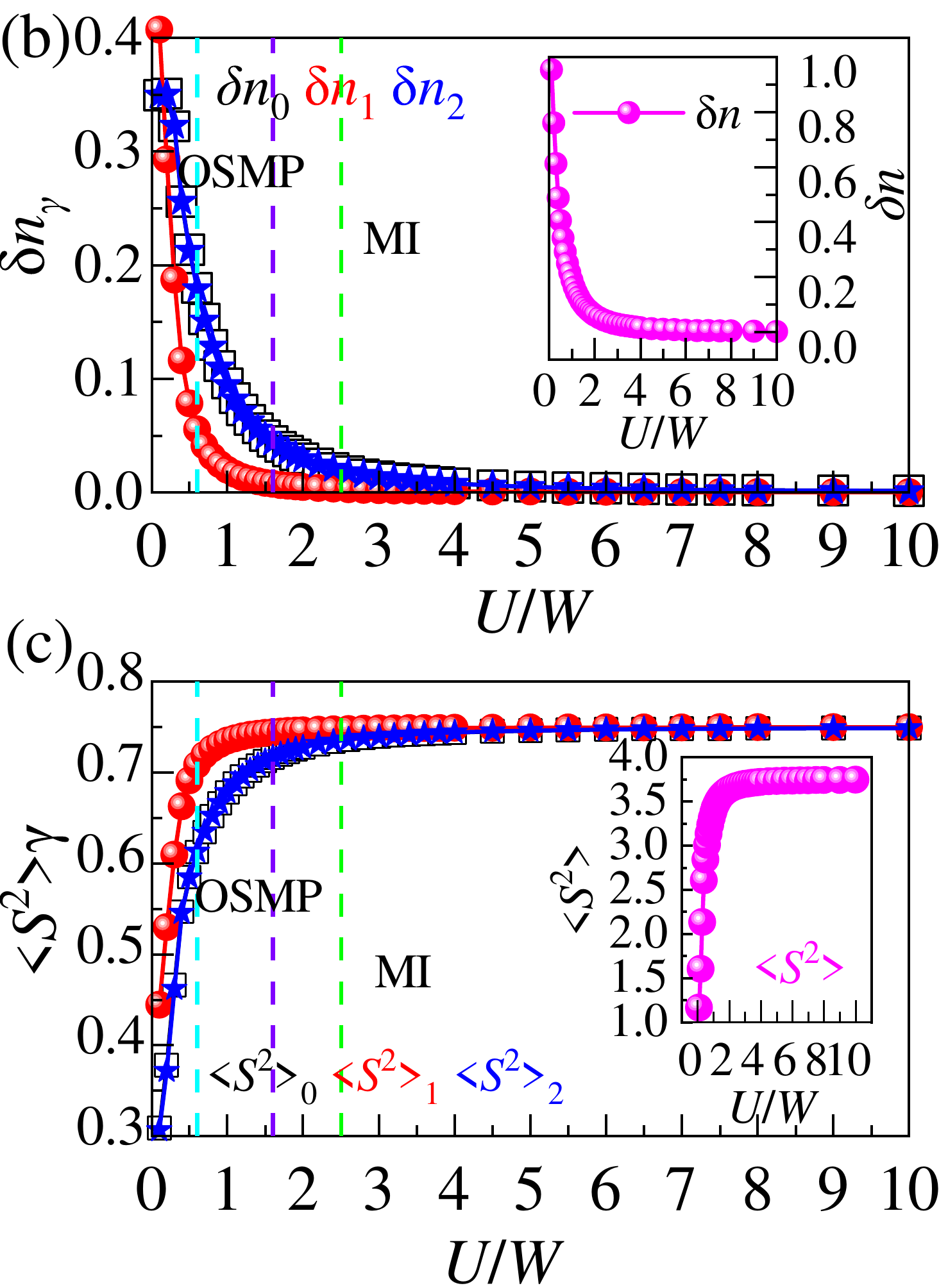}
\caption{ (a) Orbital-resolved occupation number $n_{\gamma}$ for different $U/W$, at $J_{H}/U = 0.25$. (b) Charge fluctuations $\delta{n_{\gamma}}=\frac{1}{L}\sum_{i}({\langle}n_{\gamma,i}^2\rangle-{\langle}n_{\gamma,i}{\rangle}^2)$ for different orbitals
as a function of $U/W$, at $J_H/U = 0.25$. Insert: total charge fluctuations $\delta{n}=\frac{1}{L}\sum_{i}({\langle}n^2\rangle-{\langle}n{\rangle}^2)$. (c)  The averaged value of the spin-squared $\langle{{S}}^2\rangle$$_{\gamma}$, for different orbitals, vs $U/W$, at $J_H$/$U$ = 0.25.  Insert: the averaged value of the total spin-squared $\langle{{S}}^2\rangle$. Different phases are distinguished by the dashed lines with different colors. Here, our crucial goal is to establish the dominant tendencies in the system. The phase boundary of OSMP and MI is determined by charge fluctuations of different orbitals, where the MI state is identified if there are no clear charge fluctuations. Note that those phase boundaries should be considered only as crude approximations. However, the existence of the different regions is clearly established even if the boundaries are only crude estimations. In addition, we used a lattice length $L  = 24$.}
\label{DMRG2}
\end{figure}

In addition, we calculated the site-averaged occupation number $n_{\gamma}$ as a function of $U/W$  at $J_H/U = 0.25$, as shown in Fig.~\ref{DMRG2}(a). In the small $U/W$ region, $\gamma = 0$ and $\gamma = 2$ have non-integer $n_{\gamma}$ values, indicating a metallic behavior. As $U/W$ increases, the occupation number of all orbitals converges to $1$, leading to three half-occupied states, resulting in a Mott insulator staggered AFM state.

Next, we also analyzed the charge fluctuations ($\delta{n_{\gamma}}=\frac{1}{L}\sum_{i}({\langle}n_{\gamma,i}^2\rangle-{\langle}n_{\gamma,i}{\rangle}^2$) for different orbitals, as displayed in Fig.~\ref{DMRG2}(b).
In the small $U/W$ region ($\lesssim 0.6$), the charge fluctuations for all three orbitals are considerable, indicating strong quantum fluctuations along the chains, leading to a metallic state. The
charge occupation of the $\gamma = 1$ orbital [see Fig.~\ref{DMRG2}(a)] rapidly reaches 1 with only a slight change ($\sim 2\%$) around $U/W = 0.2$, but has sizeable charge fluctuations in the small $U/W$ region.  As $U/W$ increases to
intermediate Hubbard coupling strengths, the charge fluctuations $\delta n_1$ rapidly reduce to nearly zero, resulting in localized orbital characteristics. For comparison,
the other two orbitals ($\gamma = 0$ and $\gamma = 2$)
still retain some sizeable charge fluctuations coexisting with some itinerant electron characteristics, leading to metallic orbital behavior. In this case, the system displays the OSMP in the intermediate $U/W$ regime ($\sim 0.6$ to $\sim 1.6$).

At even larger $U/W$ ($\gtrsim 2.5$), the charge fluctuations of all the orbitals are reduced to nearly zero. Thus, the charge fluctuations are virtually fully suppressed by the electronic correlations. And this system displays
insulating behavior in the strong electronic correlation region, as an AFM MI state with three half-filled orbitals ($\gamma = 0$, $\gamma = 1$, and $\gamma = 2$). As discussed in our previous study~\cite{zhang:prb21}, the AFM-OSMP to AFM-MI
is likely a ``crossover'' rather than a sharp, well-defined phase transition.

Furthermore, the calculated averaged value of the spin-squared of different orbitals also supports the above described phase transitions, as shown in Fig.~\ref{DMRG2}(c). In the small $U/W$ region,
all the spin squared $\langle{{S}}^2\rangle$$_{\gamma}$ values are unsaturated for different orbitals. In the AFM OSMP region, $\langle{{S}}^2\rangle$$_1$ is saturated ($\sim 0.75$), while $\langle{{S}}^2\rangle$$_0$
and $\langle{{S}}^2\rangle$$_2$ are still not fully converged to $0.75$, indicating the presence of magnetic fluctuations. Continuing increasing $U/W$, eventually $\langle{{S}}^2\rangle$$_{\gamma}$ of all the orbitals saturates to $0.75$,
leading to a total $\langle{{S}}^2\rangle$ reaching $3.75$ (three electrons in three orbitals),
corresponding to the one-dimensional version of an AFM MI state.

We also would like to note that the phase boundary should only be considered as crude approximations, because finite lattice size effects and the use of a limited number of states in DMRG would affect the specific boundary values of this regime change from delocalized to localized electrons. However, although the lattice size effects would affect the specific boundary values of the transitions, the existence of the different regimes is clear, even if the boundaries are only crude estimations.

Moreover, the interesting OSMP region was also reported in the Q1D iron ladder BaFe$_2$Se$_3$~\cite{mourigal:prl15,Patel:osmp,Herbrych:prl}, as well as the Fe$^{2+}$ iron chain system Ce$_2$O$_2$FeSe$_2$~\cite{LindopingOSMP1,LindopingOSMP2}.
Hence, our results clearly indicate the similarity between Fe$^{3+}$ and Fe$^{2+}$ iron low-dimensional systems, with coexisting localized and itinerant electrons.

\subsection{Orbital-selective Mott Phase}

\begin{figure*}
\centering
\includegraphics[width=0.92\textwidth]{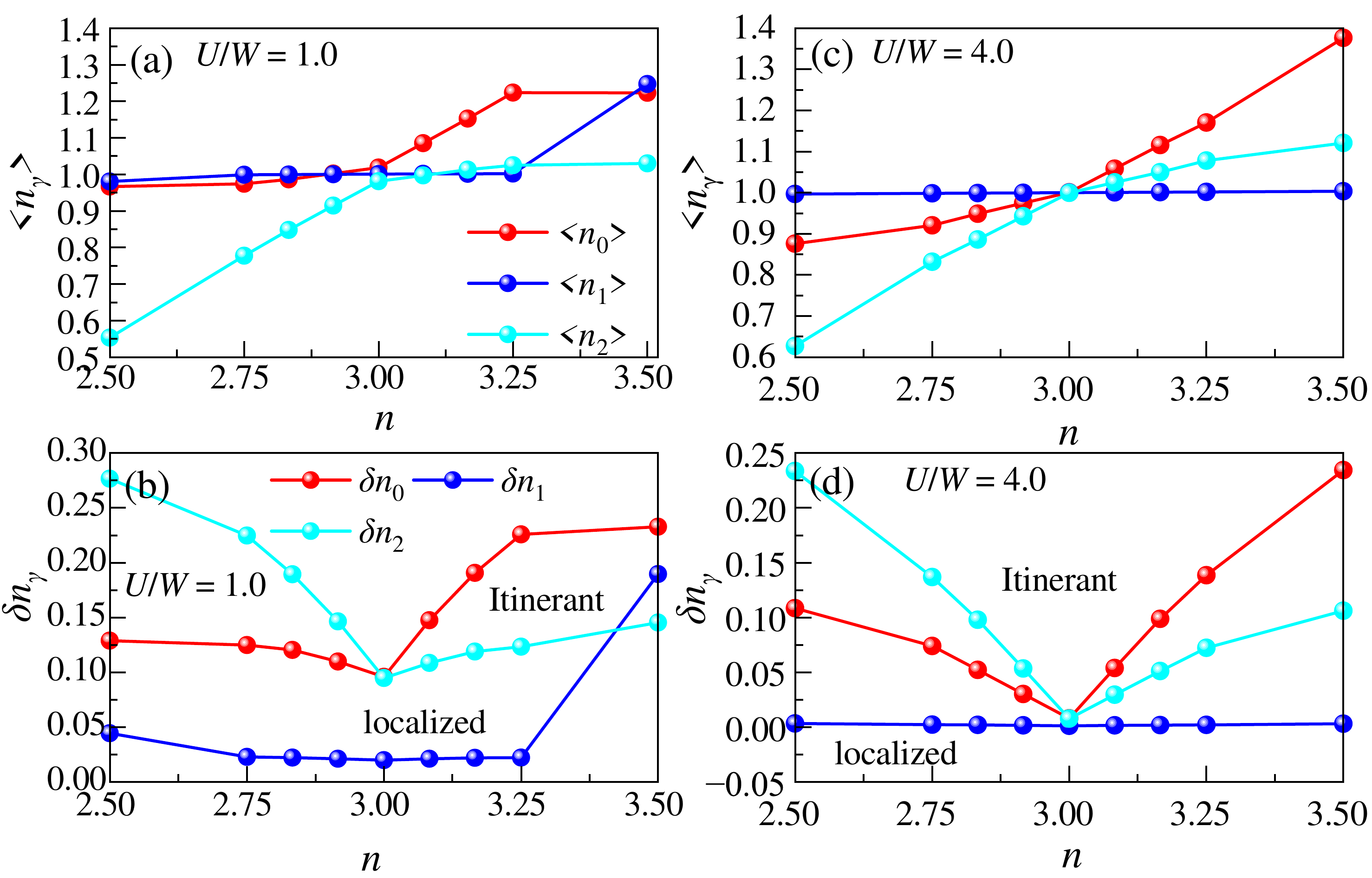}
\caption{ (a) Orbital-resolved occupation number $n_{\gamma}$ and (b) charge fluctuations $\delta{n_{\gamma}}$ of the different orbitals varying doping at $U/W = 1.0$ and $J_H/U = 0.25$. (c) Orbital-resolved occupation number $n_{\gamma}$ and (d) charge fluctuations $\delta{n_{\gamma}}$ of different orbitals varying doping at $U/W = 4.0$ and $J_H/U = 0.25$. Here, we used a chain lattice of size $L = 24$. In panels (b) and (d), ``localized'' means that the orbital is Mott-localized ($\delta{n_{\gamma}}$ $\sim 0$), and ``itinerant'' suggests the orbital is metallic with sizable charge fluctuations.}
\label{OSMP-doping}
\end{figure*}

To examine the stability of the OSMP, we also studied the occupation number and charge fluctuations for each orbital with different global filling numbers of electrons, for both intermediate and strong electronic correlations $U/W$,
respectively. At intermediate electronic correlation, the system displays the OSMP with some charge fluctuations for the ${\gamma} = 0$ and ${\gamma} = 2$ orbitals (see the $n = 3.0$ results in Figs.~\ref{OSMP-doping}(a) and (b)).
With hole doping, linear behavior was observed in the ${\gamma} = 0$ orbital with large charge fluctuations, while the ${\gamma} = 1$ orbital still keeps Mott-localized characteristics.
Furthermore, the occupation number of the ${\gamma} = 2$
orbital slightly changes containing some charge fluctuations under hole doping, as shown in Figs.~\ref{OSMP-doping}(a) and (b). For electronic doping, the ${\gamma} = 2$ orbital shows linear behavior with sizeable
charge fluctuations while the ${\gamma} = 1$ orbital remains Mott-localized in some regions of electronic doping. Thus, the OSMP can be stable in a wide range of doping at intermediate correlation.

For strong electronic correlations $U/W$, the ${\gamma} = 1$ orbital stays localized at both hole and electronic doping without charge fluctuations, as displayed in Figs.~\ref{OSMP-doping}(c) and (d). Meanwhile,
both the ${\gamma} = 0$ and ${\gamma} = 2$ orbitals display a linear behavior under doping with fractional occupation, and thus metallic character, indicating a global OSMP behavior. Hence, all these results indicate that the OSMP physics could be
stable in both the intermediate and strong electronic correlation regimes. Our DMRG results suggest that the iron 3+ chain could also be quite interesting with OSMP physics involving the coexistence of localized and itinerant electrons, as widely-discussed
in the iron 2+ system~\cite{mourigal:prl15,Patel:osmp,Craco:prb20,Emilian:qm,Benfatto:qm}. Similar to the case of BaFe$_2$Se$_3$~\cite{mourigal:prl15}, inelastic neutron scattering experiments could provide the key experimental evidence
for the static and dynamic spin correlations involving the iron ions, which deserve further experimental efforts on BaFe$_2$Se$_3$ and related Fe$^{3+}$ chains.

Here, we would like to note that our identification of the intermediate regime as an OSMP is primarily based on the orbital-selective behavior observed in the charge fluctuations, where one orbital shows strongly suppressed fluctuations while for the others they remain relatively large. Thus, one behaves as insulator while the other behave as a metal. We acknowledge that this criterion alone is not fully conclusive in identifying an OSMP, but is consistent with. A detailed dynamical spectra study for this region could provide stronger and better evidence for identifying OSMP, such as discussed in Ref.~\cite{Patel:osmp} for the iron ladder BaFe$_2$Se$_3$. This computer-time costly calculation could be considered in future work.

\subsection{Binding energy}
Considering the pressure-induced superconductivity that was reported for the Q1D iron ladders BaFe$_2$S$_3$~\cite{Takahashi:Nm} and BaFe$_2$Se$_3$~\cite{Ying:prb17}, here we also studied the binding energy of a pair of holes to
explore the possibility of pairing tendencies in the iron 3+ chain, as defined in Ref.~\cite{Patel:prb16,Patel:prb17}:

\begin{eqnarray}
{\Delta}E=E(N-2)+E(N)-2E(N-1),
\end{eqnarray}
where $E(N)$ is the ground-state energy of the $N$ electron system for the three-orbital chain model. $E(N-2)$ and $E(N-1)$ are the ground-state energy of the two-hole doped or one-hole doped cases, respectively,
corresponding to the $N-2$ and $N-1$ electron systems. If $\Delta E$ is found to be negative, it indicates possible pairing tendencies in this system, because the particles minimize
their energy by creating a bound state. If $\Delta E$ is calculated to be positive, it means the particles do not bind; thus, there are no pairing tendencies in the system. In this case and in the bulk limit, the
doped holes would become two independent particles, leading to zero binding energy.

\begin{figure}
\centering
\includegraphics[width=0.48\textwidth]{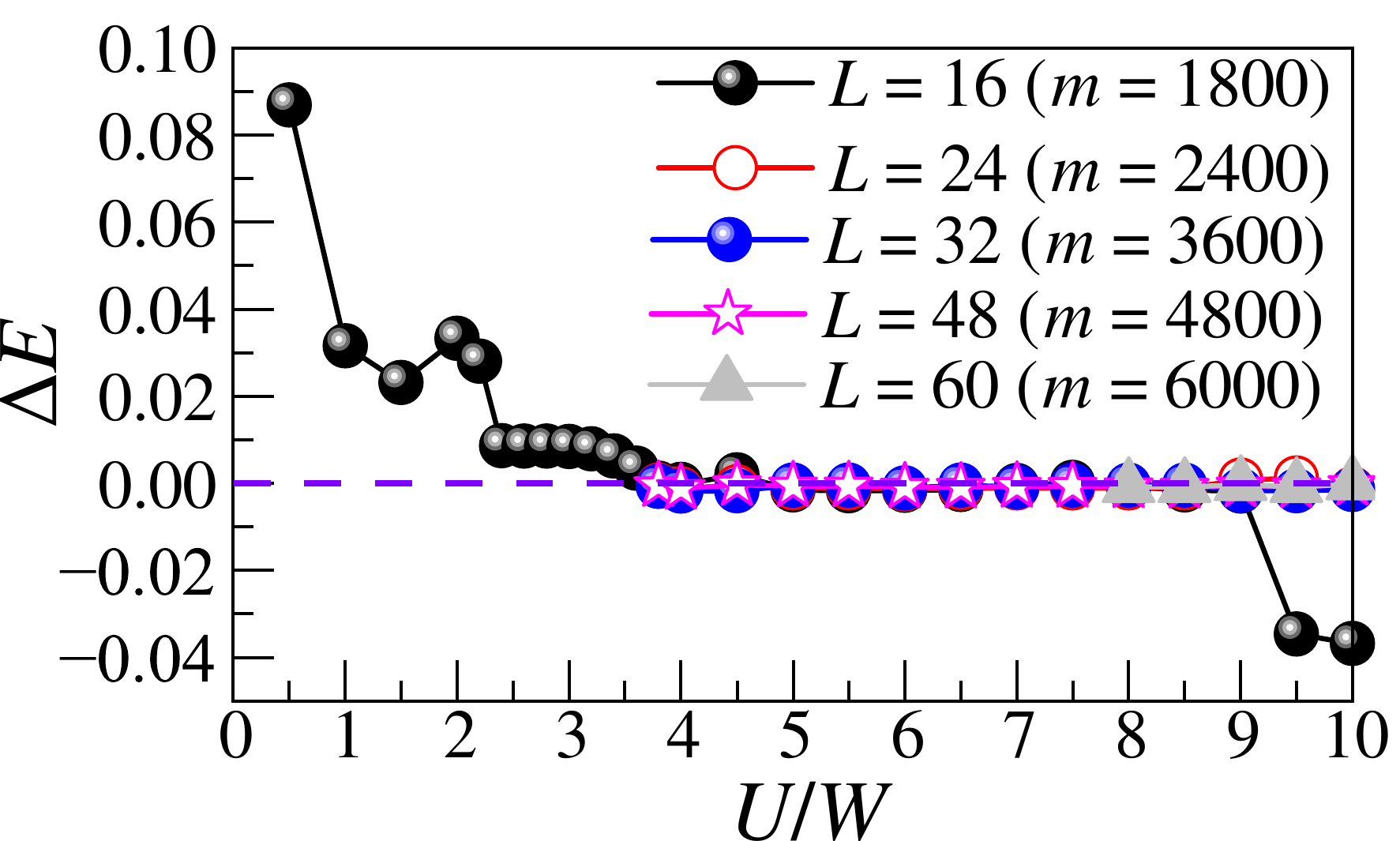}
\caption{The calculated binding energy vs electronic correlation strength $U/W$ in a chain lattice for different lengths $L$ and number of states $m$. Specifically, the maximal number of kept states were 1800, 2400, 3600, 4800, and 6000 for different lengths $L$ =
16, 24, 36, 48, and 60, respectively. The truncation error is smaller than 10$^{-6}$ everywhere except for the $L = 16$ data at $U/W \textless 3.8$. In the range between 2.0 and 3.6, it is at least snaller than 10$^{-5}$. Because the DMRG computational cost is significantly higher in the small region of $U/W$, we only focused on the large $U/W$ points when using systems with larger sizes and more states. Here, the results
were obtained at $J_H/U$ = 0.25.}
\label{Pairing}
\end{figure}

Based on the ground-state energies from DMRG calculations for the $N$ (undoped case), $N-1$ (one hole), and $N-2$ (two holes) electron systems, we obtain the binding energy for different lengths $L$ and states $m$,
as shown in Fig.~\ref{Pairing}. At electronic correlation $U/W$ $\textless 4$, the calculated values of $\Delta E$ are positive, indicating no pairing tendencies in this region for $L = 16$. In practice, the strong spin-spin correlations could obscure the emergence of pairing tendencies for a relatively small system size, such as $L = 16$. However, it would be very time-consuming to calculate spin and charge gaps for larger-size systems in this region. Indeed, we tried finite-size scaling of the pairing gap binding energy at points using larger system sizes inside the region $U/W$ $\textless 4$. The results were inconclusive and we cannot rule out a tiny negative binding energy. Those calculations are extremely computationally costly, and the results are not of high accuracy due to the computational constraints of our computing resources. Thus, more convincing finite-size scaling is postponed to further work. For comparison, using the same method, our previous efforts found a strong pairing tendency for the iron 2+ ladders at a broad electronic coupling $U/W$ regime from $\sim 1$ to $\sim 4$~\cite{Patel:prb16,Patel:prb17}. In this case, different from Fe$^{2+}$ ladders, the Fe$^{3+}$ case appears far from exhibiting pairing, and thus tendencies to superconductivity, at least at the same conditions as used in iron ladders.

Furthermore, while at large $U/W$ ($\textgreater$ 4), the calculated binding energies become slightly negative,  their magnitude remains extremely small on the order of 10$^{-4}$ eV, suggesting either the absence of pairing tendencies or,
at most, a very weak pairing effect that is likely not robust. At $L = 16$, we do see a considerable negative binding energy at very large $U/W$ as shown in Fig.~\ref{Pairing}, but it disappears as the length $L$ increases, indicating that it is likely a finite-size artifact rather than a true signature of pairing. Considering the electronic correlation of iron-based superconductor is at the intermediate region~\cite{Dai:np,Dai:Rmp,Dagotto:Rmp}, this suggests that iron 3+ chains
are unlikely to be superconducting.

\begin{figure*}
\centering
\includegraphics[width=0.88\textwidth]{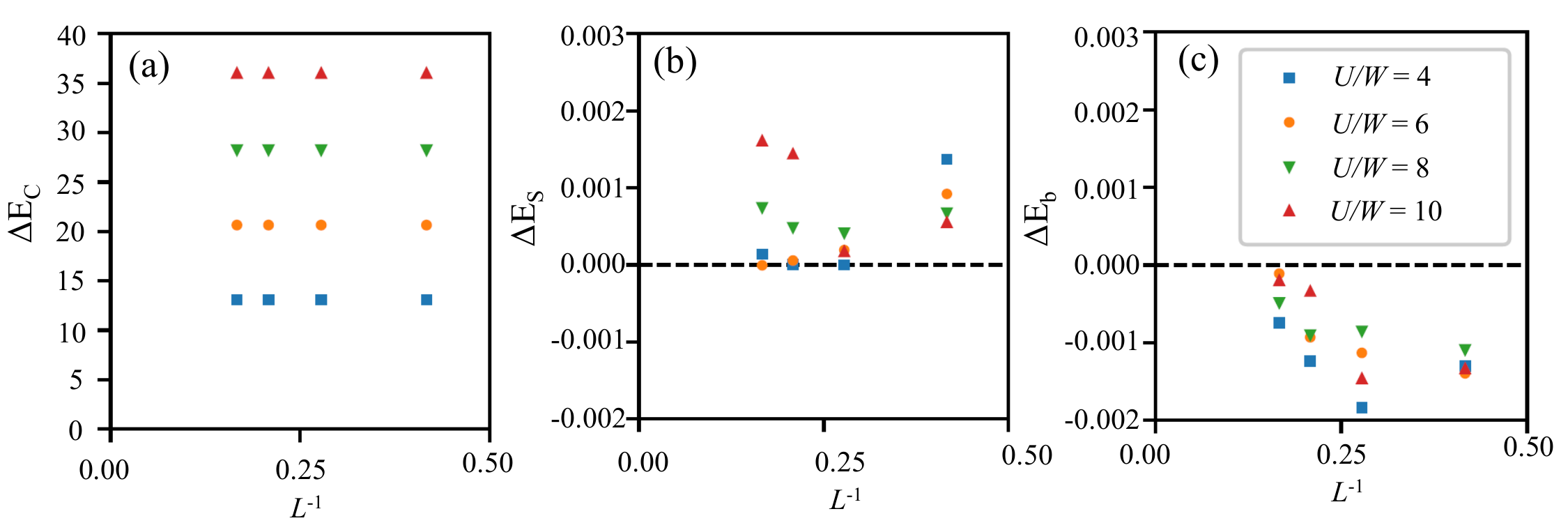}
\caption{The calculated (a) charge gap $\Delta E_c$ [$\Delta E_c=(E(N+1)+E(N-1)-2E(N)$], spin gap $\Delta E_s$ [$\Delta E_s=E(S^z = 1) - E_0(S^z = 0)$ and $S^z = (N_{\uparrow}-N_{\downarrow})$/2], and binding energy $\Delta E_b$ [$\Delta E_b=E(N-2)+E(N)-2E(N-1)$] parametric with electronic correlation strength $U/W$ using a chain with different lengths $L$. Specifically, we use $U/W$ equal to 4, 6, 8, and 10 for different lengths $L$ = 24, 36, 48, and 60, respectively. All the results were obtained at $J_H/U$ = 0.25.}
\label{Gap-L}
\end{figure*}

In addition, we also investigated spin ($\Delta E_s$) and charge ($\Delta E_c$) gaps for several large values of $U/W$ (= 4, 6, 8, 10) using different lattice sizes $L$ up to 60, as shown in Fig.~\ref{Gap-L}. We also calculated binding energies. In this regime, finite-size scaling suggests either zero or tiny $\Delta E_s$ $\textgreater 0$ and zero or tiny binding energy $\Delta E_b$  $\textless 0$. Clearly, the charge gap is larger by several orders of magnitude ($\Delta E_c \gg \Delta E_s,\quad \Delta E_c \gg \lvert \Delta E_b \rvert$). The large charge gap presumably means that charges are mostly localized and suggests an incompressible charge state. The very weak spin gap suggests that the ground state resembles more the spin arrangement of an antiferromagnet, with very low-energy spin excitations. The tiny binding energies we found suggest that pairing is very weak or there is no pairing for this system.

Recently, a high-pressure study provides support to our conclusions, where a possible superconductivity was reported but with very low $T_C^{\rm onsite}$ $\sim 2$ K above $30$ Gpa in the iron Fe$^{3+}$ chain TlFeSe$_2$~\cite{Liu:cpl}.
In addition, a possible structural phase transition was also assumed in TlFeSe$_2$ under pressure~\cite{Liu:cpl}. Then, the lattice and electronic properties could be changed substantially varying details and sample preparation and conditions.
This requires a comprehensive DFT study for this iron 3+ chain or related chains to explore the possible lattice and electronic properties in high-pressure conditions. Based on the hoppings, crystal fields, and the relevant orbitals obtained from
the high-pressure DFT study, a system model calculation can provide more physical insight for the puzzle of high-pressure superconductivity in TlFeSe$_2$, as discussed in iron ladders~\cite{Arita:prb15,Patel:qm,Pandey:prb21}. All those deserve further
theoretical investigation and discussion beyond the scope of our present paper.

\subsection{Varying the electronic density}
Recently, a half-full mechanism involving strong entanglements between doubly occupied and half-filled orbitals was proposed to explain the FM coupling along the chain direction in Fe$^{2+}$ Ce$_2$O$_2$FeSe$_2$~\cite{Lin:prl}. To better understand this mechanism
and to establish a connection between  Fe$^{3+}$ and Fe$^{2+}$ chain systems, we now examine the magnetic correlations of our present model at an electronic filling of four electrons per site, corresponding to the Fe$^{2+}$ valence.

\begin{figure}
\centering
\includegraphics[width=0.48\textwidth]{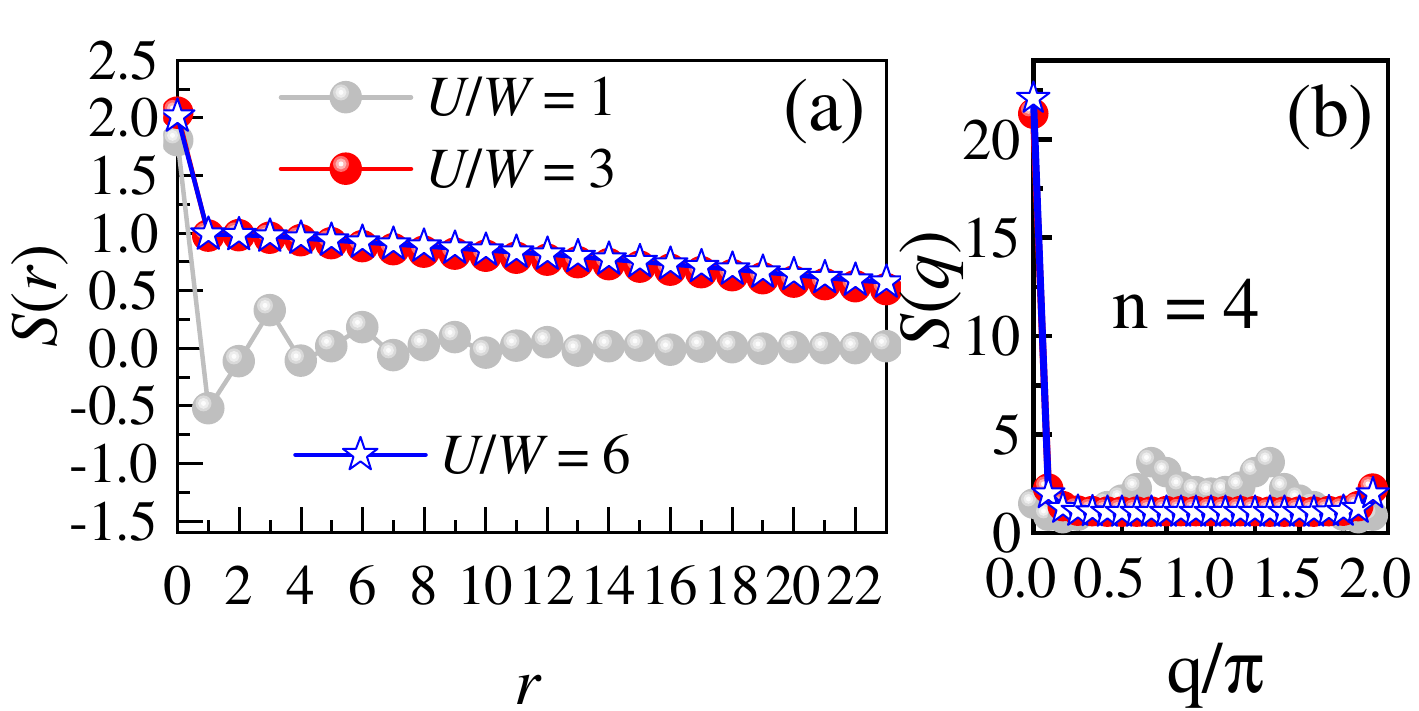}
\includegraphics[width=0.48\textwidth]{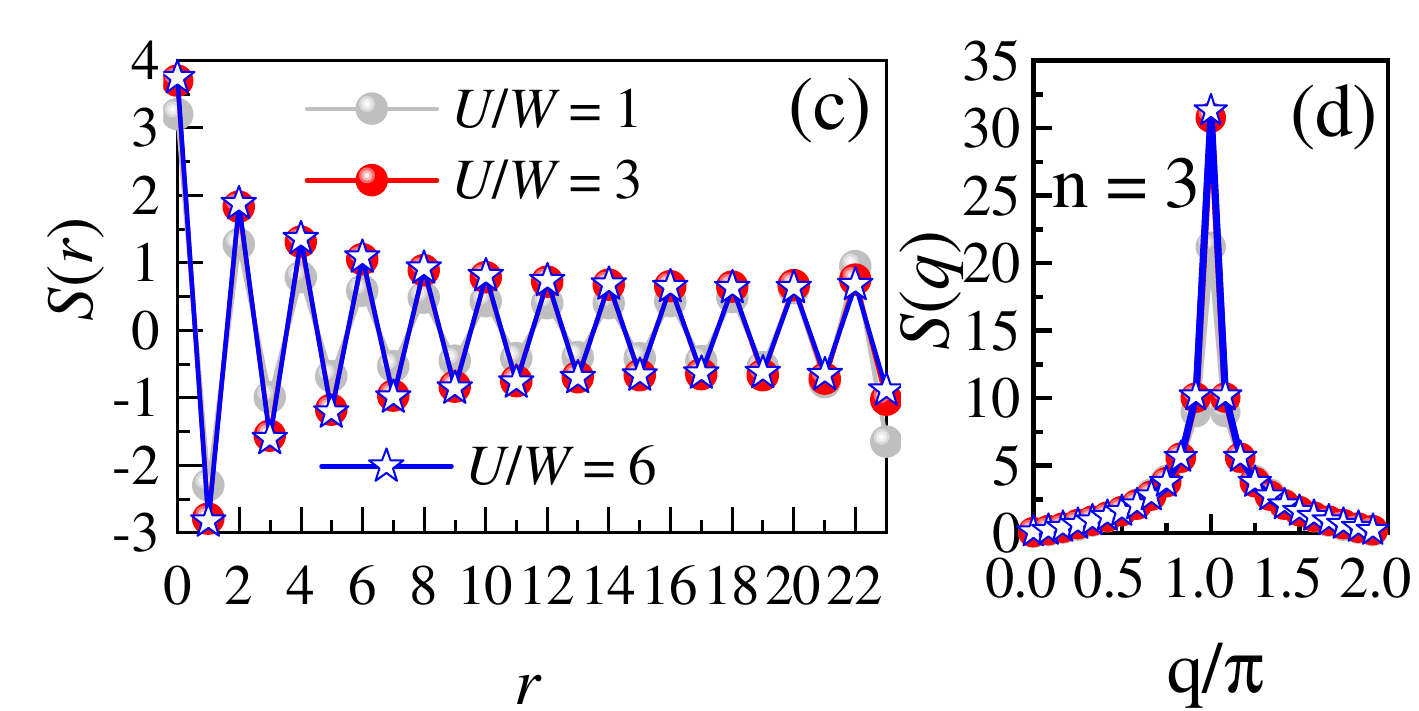}
\includegraphics[width=0.48\textwidth]{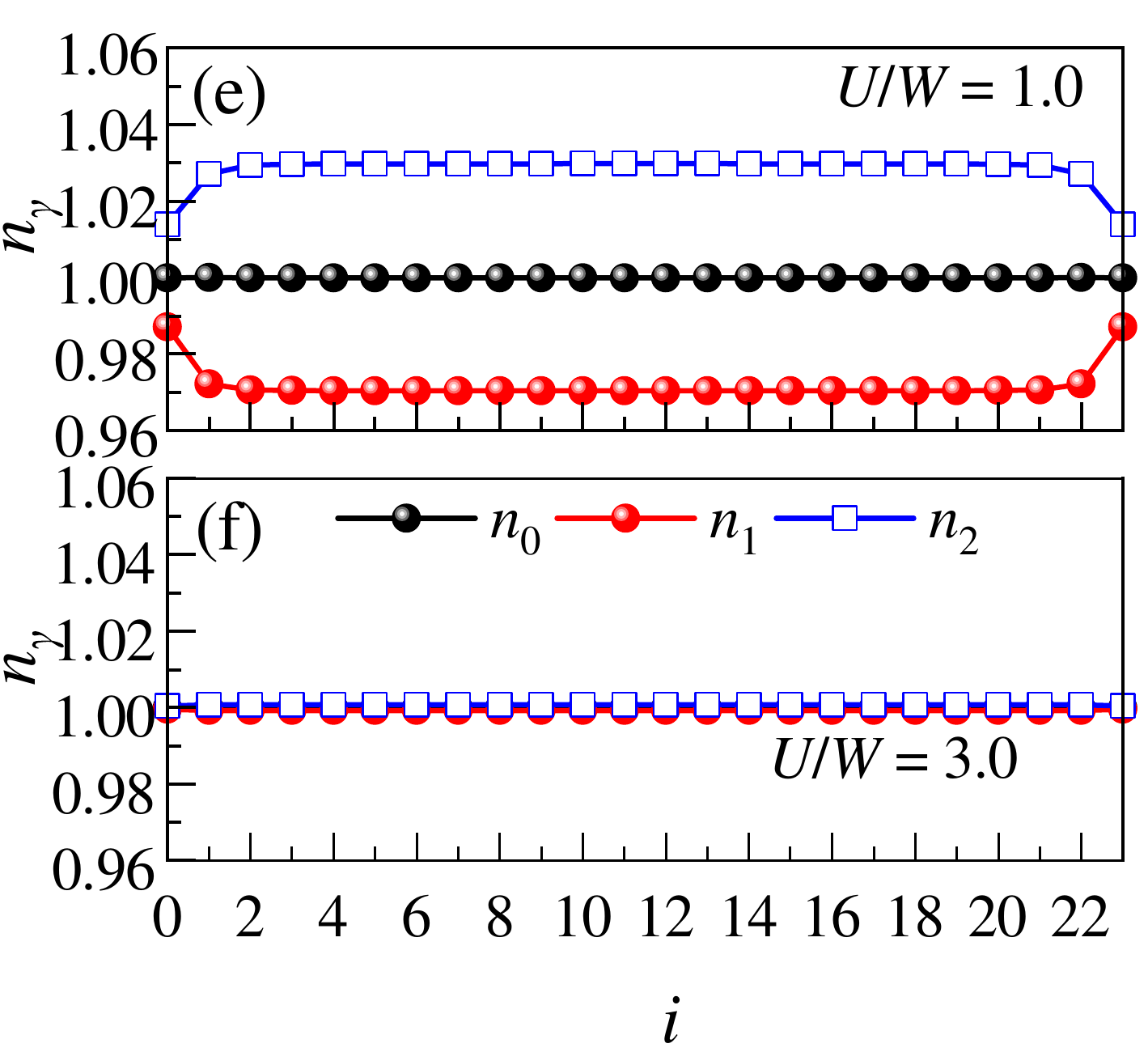}
\caption{ (a) Spin-spin correlation $S(r)$ in real space, and (b) the spin structure factor $S(q)$, as a function of $U/W$, at $J_H/U$ = 0.25, respectively, using the hopping matrix and crystal-field splitting obtained from BaFe$_2$Se$_4$. (a) and (b): here we use a chain lattice geometry with $L = 24$ and 96 electrons, corresponding to four electrons in three orbitals with 24 sites. (c) Spin-spin correlation $S(r)$ in real space, and (d) the spin structure factor $S(q)$, as a function of $U/W$, at $J_H/U$ = 0.25, respectively, using the hopping matrix and crystal-field splitting obtained from the iron 2+ chain~\cite{Lin:prl}. (c) and (d) Orbital-resolved occupation number $n_{\gamma}$ vs site $i$ for both $U/W = 1$ and $U/W = 3$, respectively, at $J_{H}/U = 0.25$. (e) to (f) Orbital-resolved occupation number $n_{\gamma}$ for different site at $U/W = 1$, and $U/W = 3$, respectively.
Here, we use a chain lattice geometry with $L = 24$ and modify the number of electrons to 72, corresponding to three electrons in three orbitals with 24 sites.}
\label{n4}
\end{figure}

At $U/W = 1$, different from the case of the three electrons in three orbitals (AFM order already forms), Fig.~\ref{n4}(a) indicates that the spin-spin correlation $S(r)$ decays as the distance $r$ increases, indicating a PM behavior.
This is caused by the competition between FM and AFM coupling, which are caused by interorbital and intraorbital couplings, respectively, at intermediate $U/W$. However, $S(r)$ also indicates the possibility of the local short-range
$\uparrow$-$\downarrow$-$\rightarrow$ order at $U/W = 1$. This is also supported by the calculated spin structure factor $S(q)$, where a small peak at $q = 2\pi/3$ was found, as displayed in Fig.~\ref{n4}(b). This magnetic phase
($\uparrow$-$\downarrow$-$\rightarrow$) is quite interesting, which deserves additional study by adjusting different parameters (electronic densities, hoppings, and crystal-field splitting) to stabilize this phase. However, we will not discuss further this possible phase, leaving the effort for future work.

As $U/W$ increases, the spin-spin correlation $S(r)$ gradually converges at larger distance, indicating FM order (see results for $U/W = 3$ or 6), as displayed in Fig.~\ref{n4}(a).  Note that the approximately linear behavior of the FM spin-spin correlation as a function of distance $r$ should not be interpreted as an intrinsic bulk scaling law. In general, FM spin-spin correlations are expected to saturate to a constant rather than exhibiting a linear dependence on distance. The observed behavior arises from the combined effects of finite system size, open boundary conditions, and the presence of strong ferromagnetic correlations in a finite chain. While larger system sizes would further reduce boundary effects, the purpose of Fig.~\ref{n4}(a) is not to extract the long-distance scaling form of the correlation function. Instead, it is intended to demonstrate the clear tendency toward ferromagnetic correlations along the chain to show the importance of electronic density in altering magnetism after introducing robust interorbital hopping. This ferromagnetic tendency is unambiguously established by the data shown in Fig.~\ref{n4}(a), which is enough to support that directly modifying the electronic density could alter the magnetic properties. The main message and conclusion discussed here does not depend on the shape of the curve. Furthermore,
the spin structure factor $S(q)$ also displays a sharp peak at $q = 0$, corresponding to the FM order (see Fig.~\ref{n4}(b)). Thus, it suggests that different magnetic coupling behaviors between Fe$^{2+}$ and Fe$^{3+}$ chains
is caused by different electronic densities, where the orbital-selective character plays an important role. Hence, the results obtained by directly modifying the electronic density also support the notion that the half-full FM mechanism~\cite{Lin:prl} is robust.

In addition, we also used a new set of hopping parameters and crystal-field splitting obtained from the iron 2+ chain~\cite{Lin:prl} and changed the electronic density to 3 in the same model (corresponding to the Fe$^{3+}$):
\begin{equation}
\begin{split}
t_{\gamma\gamma'}^{\prime} =
\begin{bmatrix}
          0.187	    &  -0.054	   &       0.020	   	       \\
          0.054	    &   0.351	   &      -0.349	   	       \\
          0.020	    &   0.349	   &      -0.433	
\end{bmatrix}.\\
\end{split}
\end{equation}
The crystal-field splitting of the three orbitals are fixed as $\Delta_{0}^{\prime} = -0.277$, $\Delta_{1}^{\prime} = -0.203$ eV and $\Delta_{2}^{\prime} = -0.720$ eV, while
the total kinetic energy bandwidth $W^{\prime}$ is $2.1$ eV.

The spin-spin correlation $S(r)$ clearly shows an AFM coupling along the chain for different $U/W$, as displayed in Fig.~\ref{n4}(c). Accordingly, the spin structure factor $S(q)$ also displays a sharp peak at $q = \pi$,
indicating AFM order (see Fig.~\ref{n4}(d)). In addition, we also studied the electronic occupation number of different orbitals $n_{\gamma}$ for $U/W = 1$ and $U/W = 3$, respectively, as shown in Figs.~\ref{n4}(e) and (f).
At $U/W = 1$, $n_0$ reaches 1 with nearly zero charge fluctuations while $n_1$ and $n_2$ are not fully close to 1, also with sizeable charge fluctuations, leading to an OSMP. However, all three orbitals are half-filled with
$n_{\gamma} = 1$ at $U/W = 1.0$, without charge fluctuations at $U/W = 3$, leading to a MI phase. Note that in this set of hoppings, the large interorbital hopping is between $\gamma_1$ and $\gamma_2$ orbitals, while the $\gamma_0$ orbital has the smallest hopping.
Hence, those qualitative similarities with other set of parameters, suggest our physical conclusion of iron 3+ chain are solid and robust.

\section{Conclusions and Discussion}
Motivated by recent interesting developments in the study of low-dimensional 1D iron systems with electronic density $n = 6$ (both two-leg ladders and chains), we investigated the iron chain system now with $n = 5$ with a
Fe$X_4$ ($X$ = S or Se) tetrahedron block. To better understand the physics behind iron chains with $n = 5$, starting from BaFe$_2$Se$_4$, here we comprehensively studied the electronic correlation effect, in a three-orbital
lattice model defined on a chain lattice, by using DMRG many-body techniques. Our results establish the similarities and differences between iron 2+ and 3+ iron chains, as well as iron ladders.

Based on the electronic structures of the non-magnetic phase of BaFe$_2$Se$_4$, the states near the Fermi level are mainly contributed by the Fe $3d$ orbitals hybridized with Se $4p$ orbitals.
Similar to the case of Ce$_2$O$_2$FeSe$_2$ with iron 2+, the five iron orbitals of BaFe$_2$Se$_4$ with iron 3+ also display entangled bonding and antibonding characteristics. Furthermore,
the interorbital hoppings (nonzero off-diagonal matrix elements) are also found to be large in BaFe$_2$Se$_4$, as reported in Ce$_2$O$_2$FeSe$_2$~\cite{Lin:prl}. Due to having a similar Fe$X_4$ ($X$ = S or Se)
chain direction, the iron 3+ chains have similar electronic structures (see Note III in the Supplemental Materials~\cite{Supplemental}), indicating similar physical properties in the Fe$^{3+}$ chain systems.

In the range of $U/W$ that we studied ($U/W \leq 10$), the spin-spin correlation $S(r)$ indicates an AFM coupling along the chain, corresponding to a sharp peak at $q = \pi$ in the spin structure factor $S(q)$. This agrees well with the
present experimental results for the $n = 5$ iron chain systems. Recently, the large {\it interorbital} hopping between half-filled and fully-occupied orbitals was proposed to explain the FM order in the $n = 6$ iron chain Ce$_2$O$_2$FeSe$_2$~\cite{Lin:prl}.
Due to a similar 1D FeSe$_2$ chain structure, we also found a larger {\it interorbital} hopping in the Fe$^{3+}$ chain case. By modifying the value of electronic density to $n = 4$, corresponding to Fe$^{2+}$, using the hopping matrix and crystal
field splitting from our case, we indeed found the FM state, supporting the novel mechanism proposed by our previous study~\cite{Lin:prl}. In addition, using the hoppings from the Ce$_2$O$_2$FeSe$_2$ material~\cite{Lin:prl}, changing the electronic density to
$n = 4$ in our model, corresponding to Fe$^{3+}$, we also obtained a strong AFM phase. This demonstrates that the magnetic structure, FM vs. AFM, is controlled by the electronic density rather than details in the band dispersion. On the other hand, note that the intra-orbital hoppings would lead to AFM tendencies, thus the competition of tendencies may lead to
many interesting phases since the electronic doping would enhance FM tendency in the $n = 5$ case. This competing aspect deserves further theoretical and experimental studies.

Our study also found the OSMP regime in the intermediate electronic correlation $U/W$ region. Because the intraorbital hopping $t_{11}$ (between $\gamma = 1$ orbitals) is much smaller than others,
the $\gamma = 1$ orbital is more easily Mott-localized at the intermediate Hubbard $U$ region while the other two orbitals still remain metallic, leading to an interesting AFM OSMP regime. The OSMP
was also reported in the iron 2+ chain~\cite{Lin:prl,LindopingOSMP1} and iron ladders~\cite{mourigal:prl15,Patel:osmp,Herbrych:prl}. This well establishes interesting similarities
between $3d^5$ and $3d^6$ iron
low-dimensional systems. In addition, the OSMP with the coexistence of localized and itinerant electrons can also be stable under doping in both intermediate and strong electronic correlations. Considering the
developments of OSMP in the low-dimensional iron systems with Fe$^{2+}$, the next interesting step is to study the dynamical excitations of this OSMP in the iron chain. We leave this task for future efforts.

Different from the strong pairing tendency in the typical electronic coupling $U/W$ regime of Fe ($\sim 1$ to $\sim 4$) in the $3d^6$ low-dimensional iron systems~\cite{Patel:prb16,Patel:prb17,Pandey:prb21},
in our study we do not obverse any robust pairing tendency in the $3d^5$ iron chain by using the same model and method. Thus, it suggests the pairing tendency, and associated superconductivity, will be difficult to rstabilize in Fe$^{3+}$ systems.

All the chains addressed here share a similar Fe$X_2$ ($X$ = S, Se, or Te) chain geometry built from the edge-sharing Fe$X_4$ tetrahedron blocks, leading to similar electronic structures. Although the specific values of hoppings and crystal-field splittings vary in those different chain materials, the key characters are the same: a large {\it interorbital} hopping, while the values of the intraorbital hopping vary greatly. Performing additional DMRG calculations for other iron chain compounds with other parameter sets will substantially increase the computational cost, and this is beyond the scope of the current work. The specific values of the phase boundary and the region range of different phases would be different for other iron chain materials because the Hamiltonian parameter sets will be different. However, we do not expect that the qualitative features of the phase diagram and the key physics will change in other iron 3+ chains. As discussed in our present work, the different magnetic properties of Fe$^{2+}$ (ferromagnetism) and Fe$^{3+}$ (antiferromagnetism) are controlled by the electronic density, related to the novel half-full interorbital mechanism~\cite{Lin:prl}. Furthermore, the physical nature of the orbital-selective Mott phase is caused by varying intraorbital hoppings of different orbitals, where the orbital with smaller hopping could be Mott-localized at a smaller $U$, while the orbital with larger hopping cannot be Mott-localized at this same $U$.

In summary, our study provides a starting point to connect the iron Fe$^{3+}$ ($n = 5$) chains with other iron Fe$^{2+}$ ($n = 6$) 1D systems. All our results indicate that the iron 3+ chain materials
have interesting physics, although pairing is probably absent.
Our work also establishes a uniform picture to understand different magnetic behaviors in Fe$^{3+}$ ($n = 5$) and Fe$^{2+}$ ($n = 6$) chains, as well as provides guidance in theory and experiments to work on the magnetism, OSMP,
and superconductivity of the iron chains.

\section{Acknowledgments}
This work was supported by the U.S. Department of Energy, Office of Science, Basic Energy Sciences, Materials Sciences and Engineering Division. Parts of the finite size scaling computations for this work were performed on the high performance computing infrastructure operated by Research Support Solutions in the Division of IT at the University of Missouri, Columbia MO (https://doi.org/10.32469/10355/97710). This manuscript has been authored by UT-Battelle, LLC, under contract DE-AC05-00OR22725 with the US Department of Energy (DOE). The US government retains and the publisher, by accepting the article for publication, acknowledges that the US government retains a nonexclusive, paid-up, irrevocable, worldwide license to publish or reproduce the published form of this manuscript, or allow others to do so, for US government purposes. DOE will provide public access to these results of federally sponsored research in accordance with the DOE Public Access Plan (https://www.energy.gov/doe-public-access-plan).

\section{Data Availability}
The dataset of the main findings of this study is openly available in the Zenodo Repository. In addition, the hopping and crystal field parameters for our DMRG calculations are available in this publication and an example input file is also available in the Supplementary Materials and Zenodo Repository. Furthermore, the ab initio DFT calculations are done with the code VASP. The DMRG code used in this study are available at \href{https://g1257.github.io/dmrgPlusPlus/}{https://g1257.github.io/dmrgPlusPlus/}.


\begin{references}
\bibitem{Takahashi:Nm} H. Takahashi, A. Sugimoto, Y. Nambu, T. Yamauchi, Y. Hirata, T. Kawakami, M. Avdeev, K. Matsubayashi, F. Du, C. Kawashima, H. Soeda, S. Nakano, Y. Uwatoko, Y. Ueda, T. J. Sato and K. Ohgushi, Pressure-induced superconductivity in the iron-based ladder material BaFe$_2$S$_3$ \href{https://doi.org/10.1038/nmat4351}{Nat. Mater. \textbf{14}, 1008 (2015).}
\bibitem{Ying:prb17} J.-J. Ying, H. C. Lei, C. Petrovic, Y.-M. Xiao and V.-V. Struzhkin, Interplay of magnetism and superconductivity in the compressed Fe-ladder compound BaFe$_2$Se$_3$ \href{https://doi.org/10.1103/PhysRevB.95.241109}{Phys. Rev. B \textbf{95}, 241109(R) (2017).}
\bibitem{Yamauchi:prl15} T. Yamauchi, Y. Hirata, Y. Ueda, and K. Ohgushi, Pressure-Induced Mott Transition Followed by a 24-K Superconducting Phase in BaFe$_2$S$_3$ \href{https://doi.org/10.1103/PhysRevLett.115.246402}{Phys. Rev. Lett. \textbf{115} 246402 (2015).}
\bibitem{Zhang:prb17} Y. Zhang, L. F. Lin, J. J. Zhang, E. Dagotto, and S. Dong, Pressure-driven phase transition from antiferromagnetic semiconductor to nonmagnetic metal in the two-leg ladders $A$Fe$_2$$X$$_3$ ($A$ = Ba, K; $X$ = S, Se) \href{https://doi.org/10.1103/PhysRevB.95.115154}{Phys. Rev. B \textbf{95}, 115154 (2017).}
\bibitem{Zheng:prb18} L. Zheng, B. A. Frandsen, C. Wu, M. Yi, S. Wu, Q. Huang, E. Bourret-Courchesne, G. Simutis, R. Khasanov, D.-X. Yao, M. Wang, and R. J. Birgeneau, Gradual enhancement of stripe-type antiferromagnetism in the spin-ladder material BaFe$_2$S$_3$ under pressure \href{https://doi.org/10.1103/PhysRevB.98.180402}{Phys. Rev. B \textbf{98}, 180402(R) (2018).}
\bibitem{Wu:prb19} S. Wu, J. Yin, T. Smart, A. Acharya, C. L. Bull, N. P. Funnell, T. R. Forrest, G. Simutis, R. Khasanov, S. K. Lewin, M. Wang, B. A. Frandsen, R. Jeanloz, and R. J. Birgeneau, Robust block magnetism in the spin ladder compound BaFe$_2$Se$_3$ under hydrostatic pressure \href{https://doi.org/10.1103/PhysRevB.100.214511}{Phys. Rev. B \textbf{100}, 214511 (2019).}
\bibitem{Zhang:prb19} Y. Zhang, L. F. Lin, A. Moreo, S. Dong, and E. Dagotto, Magnetic states of iron-based two-leg ladder tellurides \href{https://doi.org/10.1103/PhysRevB.100.184419}{Phys. Rev. B \textbf{100}, 184419 (2019).}
\bibitem{Pizarro:prm} J. M. Pizarro and E. Bascones, Strong electronic correlations and Fermi surface reconstruction in the quasi-one-dimensional iron superconductor BaFe$_2$S$_3$ \href{https://doi.org/10.1103/PhysRevMaterials.3.014801}{Phys. Rev. Mater. \textbf{3}, 014801 (2019).}
\bibitem{Sun:prb20} H. Sun, X. Li, Y. Zhou, J. Yu, B. A. Frandsen, S. Wu, Z. Xu, S. Jiang, Q. Huang, E. Bourret-Courchesne, L. Sun, J. W. Lynn, R. J. Birgeneau, and M. Wang, Nonsuperconducting electronic ground state in pressurized BaFe$_2$S$_3$ and BaFe$_2$S$_{2.5}$Se$_{0.5}$. \href{https://doi.org/10.1103/PhysRevB.101.205129}{Phys. Rev. B \textbf{101}, 205129 (2020).}
\bibitem{Zheng:cp22} W.-G. Zheng, V. Bal{\'e}dent, C. V. Colin, F. Damay, J.-P. Rueff, A. Forget, D. Colson, and P. Foury-Leylekian, Universal stripe order as a precursor of the superconducting phase in pressurized BaFe$_2$Se$_3$ Spin Ladder \href{https://doi.org/10.1038/s42005-022-00955-7}{Commun. Phys. \textbf{5}, 183 (2022).}
\bibitem{Zhang:prbblock} Y. Zhang, L. F. Lin, A. Moreo, S. Dong, and E. Dagotto, Iron telluride ladder compounds: Predicting the structural and magnetic properties of BaFe$_2$Te$_3$ \href{https://doi.org/10.1103/PhysRevB.101.144417}{Phys. Rev. B \textbf{101}, 144417 (2020)}
\bibitem{Maier:Prb22} T. A. Maier, and Elbio Dagotto, Coupled Hubbard ladders at weak coupling: Pairing and spin excitations \href{https://doi.org/10.1103/PhysRevB.105.054512}{Phys. Rev. B \textbf{105}, 054512 (2022).}
\bibitem{Caron:Prb} J. M. Caron, J. R. Neilson, D. C. Miller, A. Llobet, and T. M. McQueen, Iron displacements and magnetoelastic coupling in the antiferromagnetic spin-ladder compound BaFe$_2$Se$_3$ \href{https://doi.org/10.1103/PhysRevB.84.180409}{Phys. Rev. B \textbf{84}, 180409(R) (2011).}
\bibitem{Caron:Prb12} J. M. Caron, J. R. Neilson, D. C. Miller, K. Arpino, A.  Llobet, and T. M. McQueen, Orbital-selective magnetism in the spin-ladder iron selenides Ba$_{\rm 1-x}$K$_x$Fe$_2$Se$_3$ \href{https://doi.org/10.1103/PhysRevB.85.180405}{Phys. Rev. B \textbf{85}, 180405(R) (2012).}
\bibitem{Herbrych:prl} J. Herbrych, J. Heverhagen, N. D. Patel, G. Alvarez, M. Daghofer, A. Moreo, and E. Dagotto, Novel Magnetic Block States in Low-Dimensional Iron-Based Superconductors \href{https://doi.org/10.1103/PhysRevLett.123.027203}{Phys. Rev. Lett. \textbf{123}, 027203 (2019).}
\bibitem{Herbrych:block} J. Herbrych, J. Heverhagen, G. Alvarez, M. Daghofer, A. Moreo, and E. Dagotto, Block-spiral magnetism: An exotic type of frustrated order \href{https://doi.org/10.1073/pnas.2001141117}{Proc. Natl. Acad. Sci. USA \textbf{117}, 16226 (2020)}
\bibitem{Dong:prl14} S. Dong, J.-M. Liu, and E. Dagotto, BaFe$_2$Se$_3$: A High $T_C$ Magnetic Multiferroic with Large Ferrielectric Polarization \href{https://doi.org/10.1103/PhysRevLett.113.187204}{Phys. Rev. Lett. \textbf{113}, 187204 (2014).}
\bibitem{Aoyama:prb19} T. Aoyama, S. Imaizumi, T. Togashi, Y. Sato, K. Hashizume, Y. Nambu, Y. Hirata, M. Matsubara, and K. Ohgushi, Polar state induced by block-type lattice distortions in BaFe$_2$Se$_3$ with quasi-one-dimensional ladder structure \href{https://doi.org/10.1103/PhysRevB.99.241109}{Phys. Rev. B \textbf{99}, 241109 (R) (2019).}
\bibitem{mourigal:prl15} M. Mourigal, Shan Wu, M. B. Stone, J. R. Neilson, J. M. Caron, T. M. McQueen, and C. L. Broholm, Block Magnetic Excitations in the Orbitally Selective Mott Insulator BaFe$_2$Se$_3$ \href{https://doi.org/10.1103/PhysRevLett.115.047401}{Phys. Rev. Lett. \textbf{115}, 047401 (2015).}
\bibitem{Patel:osmp} N. D. Patel, A. Nocera, G. Alvarez, A. Moreo, S. Johnston and E. Dagotto, Fingerprints of an orbital-selective Mott phase in the block magnetic state of BaFe$_2$Se$_3$ ladders\href{https://doi.org/10.1038/s42005-019-0155-3}{Comm. Phys. \textbf{2}, 64 (2019)}
\bibitem{Craco:prb20} L. Craco,  and S. Leoni, Pressure-induced orbital-selective metal from the Mott insulator BaFe$_2$Se$_3$ \href{https://doi.org/10.1103/PhysRevB.101.245133}{Phys. Rev. B \textbf{101}, 245133 (2020).}
\bibitem{Stepanov:prl22} E. A. Stepanov, and S. Biermann, Can Orbital-Selective N\'eel Transitions Survive Strong Nonlocal Electronic Correlations? \href{https://doi.org/10.1103/PhysRevLett.132.226501}{Phys. Rev. Lett. \textbf{132}, 226501 (2024).}
\bibitem{Stepanov:prl} E. A. Stepanov,  Eliminating Orbital Selectivity from the Metal-Insulator Transition by Strong Magnetic Fluctuations \href{https://doi.org/10.1103/PhysRevLett.129.096404}{Phys. Rev. Lett. \textbf{129}, 096404 (2022).}
\bibitem{Zhang:prb18} Y. Zhang, L. F. Lin, J. J. Zhang, E. Dagotto, and S. Dong, Sequential structural and antiferromagnetic transitions in BaFe$_2$Se$_3$ under pressure \href{https://doi.org/10.1103/PhysRevB.97.045119}{Phys. Rev. B \textbf{97}, 045119 (2018).}
\bibitem{Materne:prb19} P. Materne, W. Bi, J. Zhao, M. Y. Hu, M. L. Amig\'o, S. Seiro, S. Aswartham, B. B\"uchner, and E. E. Alp, Bandwidth controlled insulator-metal transition in BaFe$_2$S$_3$: A M\"ossbauer study under pressure \href{https://doi.org/10.1103/PhysRevB.99.020505}{Phys. Rev. B \textbf{99}, 020505(R) (2019).}
\bibitem{Takubo:prb17} K. Takubo, Y. Yokoyama, H. Wadati, S. Iwasaki, T. Mizokawa, T. Boyko, R. Sutarto, F. He, K. Hashizume, S. Imaizumi, T. Aoyama, Y. Imai, and K. Ohgushi, Orbital order and fluctuations in the two-leg ladder materials BaFe$_2$$X$$_3$ ($X$ = S and Se) and CsFe$_2$Se$_3$ \href{https://doi.org/10.1103/PhysRevB.96.115157}{Phys. Rev. B \textbf{96}, 115517 (2017).}
\bibitem{Berthebaud:jssc} D. Berthebaud, O. Perez, J. Tobola, D. Pelloquin, and A. Maignan, Crystal and electronic structures of two new iron selenides: Ba$_4$Fe$_3$Se$_{10}$ and BaFe$_2$Se$_4$ \href{https://doi.org/10.1016/j.jssc.2015.07.027}{J. Solid State Chem. \textbf{230}, 293 (2015).}
\bibitem{Swinnea:jssc} J. S. Swinnea, and H.Steinfink, The crystal structure of $\beta$-BaFe$_2$S$_4$: The first member in the infinitely adaptive series Ba$_p$(Fe$_2$S$_4$)$_q$. \href{https://doi.org/10.1016/S0022-4596(80)80027-2}{J. Solid State Chem. \textbf{32}, 329 (1980).}
\bibitem{mccabe2011new} E.-E. McCabe, D.-G. Free, and J.-S. Evans, A new iron oxyselenide Ce$_2$O$_2$FeSe$_2$: synthesis and characterisation \href{https://doi.org/10.1039/c0cc03477k} {Chem. Commun. \textbf{47}, 1261-1263 (2011).}
\bibitem{mccabe2014magnetism} E. E. McCabe, C. Stock, J.-L. Bettis, M.-H., Whangbo, and J.-S.-O. Evans, Magnetism of the Fe$^{2+}$ and Ce$^{3+}$ sublattices in Ce$_2$O$_2$FeSe$_2$: A combined neutron powder diffraction, inelastic neutron scattering, and density functional study \href{https://doi.org/10.1103/PhysRevB.90.235115} {Phys. Rev. B \textbf {90}, 235115 (2014).}
\bibitem{Bronger:jssc} W. Bronger, A. Kyas, and P. M{\"u}ller, The antiferromagnetic structures of KFeS$_2$, RbFeS$_2$, KFeSe$_2$, and RbFeSe$_2$ and the correlation between magnetic moments and crystal field calculations \href{https://doi.org/10.1016/0022-4596(87)90065-X}{J. Solid State Chem. \textbf{70}, 262 (1987).}
\bibitem{Seidov:prb01} Z. Seidov, H.-A. Krug von Nidda, J. Hemberger, A. Loidl, G. Sultanov, E. Kerimova, and A. Panfilov, Magnetic susceptibility and ESR study of the covalent-chain antiferromagnets
TlFeS$_2$ and TlFeSe$_2$ \href{https://doi.org/10.1103/PhysRevB.65.014433}{Phys. Rev. B \textbf{65}, 014433 (2001).}
\bibitem{Seidov:prb16} Z. Seidov, H.-A. Krug von Nidda, V. Tsurkan, I. G. Filippova, A. G\"unther, T. P. Gavrilova, F. G. Vagizov, A. G. Kiiamov, L. R. Tagirov, and A. Loidl, Magnetic properties of the covalent chain antiferromagnet RbFeSe$_2$ \href{https://doi.org/10.1103/PhysRevB.94.134414}{Phys. Rev. B \textbf{94}, 134414 (2016).}
\bibitem{Li:prb21} L. Li, L. Zheng, B. A. Frandsen, A. D. Christianson, D.-X. Yao, M. Wang, and R. J. Birgenea, Spin dynamics of the spin-chain antiferromagnet RbFeS$_2$ \href{https://doi.org/10.1103/PhysRevB.104.224419}{Phys. Rev. B \textbf{104}, 224419 (2021).}
\bibitem{momma2011vesta} K. Momma, and F. Izumi, Vesta 3 for three-dimensional visualization of crystal, volumetric and morphology data \href{https://doi.org/10.1107/S0021889811038970} {J. Appl. Crystallogr. \textbf {44}, 1272 (2011).}
\bibitem{Liu:prb20} X. Liu, K. M. Taddei, S. Li, W. Liu, N. Dhale, R. Kadado, D. Berman, C. D. Cruz, and B. Lv, Canted antiferromagnetism in the quasi-one-dimensional iron chalcogenide BaFe$_2$Se$_4$ \href{https://doi.org/10.1103/PhysRevB.102.180403} {Phys. Rev. B \textbf {102}, 180403(R) (2020).}
\bibitem{Lin:prl} L. F. Lin, Y. Zhang, G. Alvarez, A. Moreo, and E. Dagotto, Origin of Insulating Ferromagnetism in Iron Oxychalcogenide Ce$_2$O$_2$FeSe$_2$, \href{https://doi.org/10.1103/PhysRevLett.127.077204}{Phys. Rev. Lett. \textbf{127}, 077204 (2021).}
\bibitem{LindopingOSMP2} L. F. Lin, Y. Zhang, G. Alvarez, M, A, McGuire, A. F. May, A. Moreo, and E. Dagotto, Stability of the interorbital-hopping mechanism for ferromagnetism in multi-orbital Hubbard models \href{https://doi.org/10.1038/s42005-023-01314-w}{Commun. Phys. \textbf{6}, 199 (2023).}
\bibitem{LindopingOSMP1} L. F. Lin, Y. Zhang, G. Alvarez, A. Moreo, Jacek Herbrych, and E. Dagotto, Prediction of orbital-selective Mott phases and block magnetic states in the quasi-one-dimensional iron chain Ce$_2$O$_2$FeSe$_2$ under hole and electron doping
 \href{https://doi.org/10.1103/PhysRevB.105.075119}{Phys. Rev. B \textbf{105}, 075119 (2022).}
\bibitem{CsFeSe2} P. St{\"u}ble, and C. R{\"o}hr, Cs[FeSe$_2$], Cs$_3$[FeSe$_2$]$_2$, and Cs$_7$[Fe$_4$Se$_8$]: Missing Links of Known Chalcogenido Ferrate Series \href{https://doi.org/10.1002/zaac.201700263} {Z. Anorg. Allg. Chem. \textbf{643}, 1462 (2017).}
\bibitem{Klepp:1979} K. Klepp, and H. Boller, Die Kristallstruktur von TlFeSe$_2$ und TlFeS$_2$ \href{https://doi.org/10.1007/BF00910952} {Monatshefte f{\"u}r Chemie/Chemical Monthly. \textbf{110}, 1045 (1979).}
\bibitem{Ho:1985} Y. Ho, F. Nishi, C. F. Majkrzak,  and L. Passell, Low Temperature Powder Neutron Diffraction Studies of CsFeS$_2$ \href{https://doi.org/10.1143/JPSJ.54.348} {J. Phys. Soc. Jpn. \textbf{54}, 348 (1985).}
\bibitem{Kresse:Prb} G. Kresse and J. Hafner, Ab initio molecular dynamics for liquid metals \href{https://doi.org/10.1103/PhysRevB.47.558}{Phys. Rev. B \textbf{47}, 558(R) (1993).}
\bibitem{Kresse:Prb96} G.~Kresse and J.~Furthm\"{u}ller, Efficient iterative schemes for ab initio total-energy calculations using a plane-wave basis set \href{https://doi.org/10.1103/PhysRevB.54.11169}{Phys. Rev. B \textbf{54}, 11169 (1996).}
\bibitem{Blochl:Prb} P. E. Bl\"{o}chl, Projector augmented-wave method \href{https://doi.org/10.1103/PhysRevB.50.17953}{Phys. Rev. B \textbf{50}, 17953 (1994).}
\bibitem{Perdew:Prl} J. P. Perdew, K. Burke, and M. Ernzerhof, Generalized Gradient Approximation Made Simple \href{https://doi.org/10.1103/PhysRevLett.77.3865}{Phys. Rev. Lett. \textbf{77}, 3865 (1996).}
\bibitem{mostofi:cpc} A. A. Mostofi, J. R. Yates, Y. S. Lee, I. Souza, D. Vanderbilt, and N. Marzari, wannier90: A tool for obtaining maximally-localised Wannier functions \href{https://doi.org/10.1016/j.cpc.2007.11.016}{Comput. Phys. Commun. \textbf{178}, 685 (2007).}
\bibitem{white:prl} S.-R. White, Density matrix formulation for quantum renormalization groups \href{https://doi.org/10.1103/PhysRevLett.69.2863}{Phys. Rev. Lett. \textbf {69}, 2863 (1992).}
\bibitem{white:prb} S.-R. White, Density-matrix algorithms for quantum renormalization groups \href{https://doi.org/10.1103/PhysRevB.48.10345}{Phys. Rev. B \textbf{48}, 10345  (1993).}
\bibitem{schollwock2005density} U. Schollw{\"o}ck, The density-matrix renormalization group \href{https://doi.org/10.1103/RevModPhys.77.259}{Rev. Mod. Phys. \textbf{77}, 259 (2005).}
\bibitem{alvarez2009density} G. Alvarez, The density matrix renormalization group for strongly correlated electron systems: A generic implementation \href{https://doi.org/10.1016/j.cpc.2009.02.016}{Comput. Phys. Commun. \textbf{180} 1572 (2009).}
\bibitem{Rincon:prl14} J. Rincon, A. Moreo, G. Alvarez, and E. Dagotto, Exotic Magnetic Order in the Orbital-Selective Mott Regime of Multi orbital Systems \href{https://doi.org/10.1103/PhysRevLett.112.106405}{Phys. Rev. Lett. \textbf{112}, 106405 (2014).}
\bibitem {zhang:prb21} Y. Zhang, L.-F. Lin, G. Alvarez, A. Moreo, and E. Dagotto, Magnetic states of quasi-one-dimensional iron chalcogenide Ba$_2$FeS$_3$ \href {https://doi.org/10.1103/PhysRevB.104.125122}{Phys. Rev. B \textbf{104}, 125122 (2021)}.
\bibitem{Supplemental}{See the Supplemental Material for additional methodological details and results, including the construction used to reproduce the DMRG results, Wannier function fitting, DFT calculations of other $n = 5$ iron chains, and extended DMRG data.}
\bibitem{Dagotto:Rmp} E. Dagotto, Colloquium: The unexpected properties of alkali metal iron selenide superconductors \href{https://doi.org/10.1103/RevModPhys.85.849}{Rev. Mod. Phys. \textbf{85}, 849 (2013).}
\bibitem{Dai:Rmp} P. Dai, Antiferromagnetic order and spin dynamics in iron-based superconductors \href{https://doi.org/10.1103/RevModPhys.87.855}{Rev. Mod. Phys. \textbf{87}, 855 (2015).}
\bibitem{Ootsuki:prb15} D. Ootsuki, N. L. Saini, F. Du, Y. Hirata, K. Ohgushi, Y. Ueda, and T. Mizokawa, Coexistence of localized and itinerant electrons in BaFe$_2$$X_3$ ($X$ = S and Se) revealed by photoemission spectroscopy \href{https://doi.org/10.1103/PhysRevB.91.014505} {Phys. Rev. B \textbf{91}, 014505 (2015).}
\bibitem{Dai:np} P. Dai, J. P. Hu, and E. Dagotto, Magnetism and its microscopic origin in iron-based high-temperature superconductors \href{https://doi.org/10.1038/nphys2438}{Nature Phys. \textbf{8}, 709 (2012).}
\bibitem{Herbrych:nc18} J. Herbrych, N. Kaushal, A. Nocera, G. Alvarez, A. Moreo, and E. Dagotto, Spin dynamics of the block orbital-selective Mott phase \href{https://doi.org/10.1038/s41467-018-06181-6}{Nat. Commun. \textbf{9}, 3736  (2018).}
\bibitem{Daghofer:prb10} M. Daghofer, A. Nicholson, A. Moreo, and E. Dagotto, Three orbital model for the iron-based superconductors \href{https://doi.org/10.1103/PhysRevB.81.014511}{Phys. Rev. B \textbf{81}, 014511 (2010).}
\bibitem{Luo:prb10} Q. Luo, G. Martins, D.-X. Yao, M. Daghofer, R. Yu, A. Moreo, and E. Dagotto, Neutron and ARPES constraints on the couplings of the multiorbital Hubbard model for the iron pnictides \href{https://doi.org/10.1103/PhysRevB.82.104508}{Phys. Rev. B \textbf{82}, 104508 (2010).}
\bibitem{Gao:prl} F. Gao, N. Dhale, L.-F. Lin, K. M. Taddei, Y. Zhang, C. Dela Cruz, E. Dagotto, and B. Lv, Block-type antiferromagnetism in single chain quasi-one-dimensional K$_3$Fe$_2$Se$_4$, to be published soon.
\bibitem{Spin-spin}Regarding the choice of reference site, we just followed our previous DMRG studies on the one-dimensional chain or ladder systems by setting the leftmost edge as the reference site to calculate the spin-spin correlation from other sites $j$.  However, the dominant underlying magnetic tendency remains unchanged if other conventions are used, as discussed in Supplementary Materials IV.
\bibitem{Emilian:qm} M. N. Emilian, R. Yu, and Q. Si, Orbital-selective pairing and superconductivity in iron selenides \href{https://doi.org/10.1038/s41535-017-0027-6}{npj Quantum Mater. \textbf {2}, 24 (2017).}
\bibitem{Benfatto:qm} L. Benfatto, B. Valenzuela, and L. Fanfarillo, Nematic Pairing from Orbital Selective Spin Fluctuations in FeSe \href{https://doi.org/10.1038/s41535-018-0129-9} {npj Quantum Mater. \textbf {3}, 56 (2018).}
\bibitem{Patel:prb16} N. D. Patel, A. Nocera, G. Alvarez, R. Arita, A. Moreo, and Elbio Dagotto, Magnetic properties and pairing tendencies of the iron-based superconducting ladder BaFe$_2$S$_3$: Combined ab initio and density matrix renormalization group study \href {https://doi.org/10.1103/PhysRevB.94.075119} {Phys. Rev. B \textbf {94}, 075119 (2016).}
\bibitem{Patel:prb17} N. D. Patel, A. Nocera, A. Moreo,  and E. Dagotto,  Pairing tendencies in a two-orbital Hubbard model in one dimension \href{https://doi.org/10.1103/PhysRevB.96.024520} {Phys. Rev. B \textbf {96}, 024520 (2017).}
\bibitem{Liu:cpl} Z.-Y. Liu, Q.-X. Dong, P.-F. Shan, Y.-Y. Wang, J.-H. Dai, R. Jana, K.-Y. Chen, J.-P. Sun, B.-S. Wang, X.-H. Yu, G.-T. Liu, Y. Uwatoko, Y. Sui, H.-X. Yang, G.-F. Chen, and J.-G. Cheng, Pressure-Induced Metallization and Structural Phase Transition in the Quasi-One-Dimensional TlFeSe$_2$ \href{https://doi.org/10.1088/0256-307X/37/4/047102}{Chin. Phys. Lett. \textbf {37}, 047102 (2020).}
\bibitem{Arita:prb15} R. Arita, H. Ikeda, S. Sakai, and M.-T. Suzuki,  Ab initio downfolding study of the iron-based ladder superconductor BaFe$_2$S$_3$ \href{https://doi.org/10.1103/PhysRevB.92.054515} {Phys. Rev. B \textbf {92}, 054515 (2015)}.
\bibitem{Patel:qm} N. D. Patel, N. Kaushal, A. Nocera, G. Alvarez, and E. Dagotto, Emergence of superconductivity in doped multiorbital Hubbard chains \href{https://doi.org/10.1038/s41535-020-0228-2} {npj Quantum Mater. \textbf {5}, 27 (2020)}.
\bibitem{Pandey:prb21} B. Pandey, Y. Zhang, N. Kaushal, R. Soni, L.-F. Lin, W.-J. Hu, G. Alvarez, and E. Dagotto, Intertwined charge, spin, and pairing orders in doped iron ladders \href{https://doi.org/10.1103/PhysRevB.103.045115}{Phys. Rev. B \textbf {103}, 045115 (2021).}
\end{references}
\end{document}